\newif\ifsingle
\singletrue 

\newif\ifFullVersion

\ifsingle
\documentclass[12pt,draftclsnofoot, onecolumn]{IEEEtran}		
\else		
\documentclass[10pt,final, twocolumn]{IEEEtran}
\usepackage[a4paper, total={7.3in, 10.4in}]{geometry}
\fi

\usepackage{times}
\usepackage{float}
\usepackage{amsmath,dsfont}
\usepackage{amssymb,amsthm}
\usepackage{epsfig,verbatim}
\usepackage{setspace}
\usepackage{color}
\usepackage{cite}
\usepackage{epstopdf}
\usepackage{graphics}
\usepackage{accents}
\usepackage{acronym}
\usepackage{enumitem}
\usepackage[bookmarks,colorlinks]{hyperref}
\usepackage[linesnumbered,lined,boxed,commentsnumbered]{algorithm2e}
\usepackage{algorithmic}
\usepackage{subcaption}

\newtheorem*{theorem*}{Theorem}

\newcommand{\myVec}[1]{{\boldsymbol{#1}}}
\newcommand{\myMat}[1]{\mathsf{#1}}
\newcommand{\mySet}[1]{\mathcal{#1}}

\newcommand{\Tnom}{T_{\mbox{\scriptsize nom}}}

\newcommand{\Wmat}{\mathsf{W}}

\newcommand{\mN}{\mathcal{N}}
\newcommand{\mI}{\mathcal{I}}
\newcommand{\mZ}{\mathcal{Z}}

\newcommand{\mX}{\mathcal{X}}

\newcommand{\xvec}{\mathbf{x}}
\newcommand{\bvec}{\mathbf{b}}

\newcommand{\yvec}{\mathbf{y}}

\newcommand{\black}{\color{black}}

\newif\ifcomments
\definecolor{CmtColor}{rgb}{0,0.6,1}

\long\def\symbolfootnote[#1]#2{\begingroup\def\thefootnote{\fnsymbol{footnote}}\footnote[#1]{#2}\endgroup}

\newcommand{\pt}{P_t}
\newcommand{\pr}{P_{i,j}}
\newcommand{\stampi}{\phi_i}
\newcommand{\stampj}{\phi_j}

\newcommand{\dij}{d_{i,j}}
\newcommand{\qij}{q_{i,j}}
\newcommand{\aij}{\alpha_{i,j}}

\acrodef{std}[STD]{standard deviation}
\acrodef{los}[LOS]{line-of-sight}
\acrodef{gps}[GPS]{Global Positioning System}
\acrodef{nn}[NN]{neural network}
\acrodef{dnn}[DNN]{deep neural network}
\acrodef{wsn}[WSN]{wireless sensor network}
\acrodef{rv}[RV]{random variable}
\acrodef{dt}[DT]{discrete-time} 
\acrodef{ct}[CT]{continuous-time} 
\acrodef{wscs}[WSCS]{wide-sense cyclostationary}
\acrodef{wsacs}[WSACS]{wide-sense almost cyclostationary}
\acrodef{ofdm}[OFDM]{orthogonal frequency division multiplexing}
\acrodef{tdma}[TDMA]{time division multiple access}
\acrodef{cdma}[CDMA]{code division multiple access}
\acrodef{noma}[NOMA]{non-orthogonal multiple access}
\acrodef{asmcgn}[AS-MCGNC]{asynchronously-sampled memoryless cyclostationary Gaussian noise channel}
\acrodef{cmt}[CMT]{continuous mapping theorem}
\acrodef{iot}[IOT]{internet of things}
\acrodef{lti}[LTI]{linear, time-invariant}
\acrodef{iiot}[IIOT]{industrial internet of things}
\acrodef{ptp}[PTP]{Precision Time Protocol}
\acrodef{ntp}[NTP]{Network Time Protocol}
\acrodef{gptp}[gPTP]{generalized Precision Time Protocol}
\acrodef{ptcp}[PTCP]{Precision Transparent Clock Protocol}
\acrodef{ptcpkf}[PTCP-KF]{Precision Transparent Clock Protocol with Kalman filter}
\acrodef{dl}[DL]{deep learning}
\acrodef{pco}[PCO]{pulse-coupled oscillator}
\acrodef{ftsp}[FTSP]{flooding time synchronization protocol}
\acrodef{pll}[PLL]{phase locked loop}
\acrodef{drl}[DRL]{deep reinforcement learning}
\acrodef{vcc}[VCC]{voltage controlled clock}
\acrodef{td}[TD]{time difference detector}
\acrodef{npd}[NPD]{normalized phase difference}
\acrodef{pdd}[PDD]{phase difference detector}
\acrodef{gd}[GD]{gradient descent}
\acrodef{pd}[PD]{phase discriminator}
\acrodef{spd}[SPD]{signal power detector}
\acrodef{mlp}[MLP]{multi-layered perceptron}
\acrodef{dasa}[DASA]{\ac{dnn}-aided synchronization algorithm}
\acrodef{mbgd}[MB-SGD]{mini-batch stochastic gradient descent}

\newcommand{\Nir}[1]{\textcolor{red}{\footnotesize{\textsf {[Nir: #1]}}}}
\newcommand{\Emk}[1]{\textcolor{blue}{\footnotesize{\textsf {[Emeka: #1]}}}}

\IEEEoverridecommandlockouts
\title{Deep-Learning-Aided Distributed Clock Synchronization for Wireless Networks
}

\author{
	\IEEEauthorblockN{ Emeka Abakasanga, Nir Shlezinger, and Ron Dabora \\
	}
	
	\thanks{
		E. Abakasanga, N. Shlezinger and R. Dabora  are with the School of ECE, Ben-Gurion University,  Israel (e-mail:  abakasan@post.bgu.ac.il; nirshl@bgu.ac.il;  daborona@bgu.ac.il).
		 This work was supported by the Israel Science Foundation under Grant 584/20 and by the Israeli Ministry of Economy via the 5G-WIN Consortium.}
	
	\vspace{-0.5cm}
	
}

\begin{document}
	
	\maketitle
	\pagestyle{plain}
	\thispagestyle{plain}
	\vspace{-1.2cm}
	\begin{abstract}
	The proliferation of wireless communications networks 
	over the past decades,
	combined with the scarcity of the wireless spectrum, have motivated a significant effort towards increasing the throughput of wireless  networks. One of the major factors which limits the throughput in wireless communications networks is the accuracy of the time synchronization between the nodes in the network, as a higher throughput requires higher synchronization accuracy.
	Existing time synchronization schemes, and particularly, methods based on pulse-coupled oscillators (PCOs), which are the focus of the current work, have the advantage of simple implementation and achieve high accuracy when the nodes are closely located, yet tend to achieve poor synchronization performance for distant nodes. In this study, we propose a robust PCO-based time synchronization algorithm which retains the simple structure of existing approaches while operating reliably and converging quickly for both distant and closely located nodes. This is achieved by augmenting PCO-based synchronization with deep learning tools that are trainable in a distributed manner, thus allowing the nodes to train their  neural network component of the synchronization algorithm without requiring additional exchange of information or central coordination.  
	The numerical results 
	show that  our proposed deep learning-aided scheme is notably  robust to propagation delays  resulting from deployments over large areas,  and to relative clock frequency offsets. It is also shown that the proposed approach  
	rapidly attains full (i.e., clock frequency and phase) synchronization for all nodes in the wireless network, while the classic model-based implementation does not. 
	
	\end{abstract}
	
	\vspace{-0.3cm}
	\section{Introduction}
	\label{sec:Intro}
	\vspace{-0.1cm}
  Time synchronization stands as a primary precondition for many applications,
  making it a broad and crucial field of research. 
    In particular, time synchronization is critical for the successful operation of wireless communications networks relying on \ac{tdma} to facilitate resource sharing. 
  With \ac{tdma}, each connected device in the network is allocated a dedicated time slot for transmission. Therefore, time synchronization among the devices 
  is essential to  ensure that there are no collisions, facilitating spectral efficiency maximization \cite{perry2014fastpass, geng2018exploiting}. 
  One example for the importance of clock synchronization in \ac{tdma}-based networks is the deployment of  \acp{wsn} in hazardous and/or secluded environments: In such scenarios, it may be impractical to recharge or replace the battery at the sensor nodes \cite{akhlaq2013rtsp}. To save power, an accurately synchronized \ac{tdma} scheme can be applied to \acp{wsn} such that the nodes are in sleep mode except during the \ac{tdma} slots in which they transmit \cite{wu2010clock}, \cite{raza2017critical}. 
  
  Synchronization can be achieved via various approaches, which can be classified as either local synchronization (involving the use of clustered nodes) or global synchronization (where all nodes are synchronized to a global clock).
  In the context of ad-hoc wireless networks, it is typically preferable for the nodes to synchronize in a distributed manner, such that the nodes in the network obtain and maintain the same network clock time independently, without requiring  direct communications with a global synchronization  device \cite{simeone2008distributed}. 
  Thus, distributed synchronization is more robust to jamming, and can be applied in scenarios in which commonly used global clocks, such as the \ac{gps}, are unavailable, e.g., in underground setups. 
  Traditional distributed time synchronization algorithms require periodic transmission and reception of time information, which is commonly implemented via packets containing a time-stamp data, exchanged between the coupled nodes \cite{sivrikaya2004time}. Packet-based synchronization has been broadly studied for wired and wireless network, \cite{karthik2020recent}, with proposed protocols including the \acl{ftsp} \cite{maroti2004flooding}, \acl{ptp} \cite{ptpprotocol}, \acl{ntp} \cite{mills2010network}, \acl{gptp} \cite{gptpIEEE}, and \acl{ptcp} \cite[Sec. 3.5]{pigan2008automating}. These approaches differ in the the way the time-stamp information is encoded, conveyed and processed across the nodes.
  The major drawbacks of packet-based coupling  are the inherent unknown delays in packet formation, queuing at the MAC layer, and packet processing at the receiver. These delays could potentially make the received time stamp carried by the packet outdated after processing is completed. Another significant drawback is the high energy consumption due to the associated processing \cite{simeone2008distributed}. 
  
  An alternative approach to packet-based synchronization, which offers lower energy consumption and simpler processing, is to utilize the broadcasting nature of the wireless medium for synchronization at the physical-layer. In this approach, the time information corresponds to the time at which the waveform transmitted by a node is received at each of the other nodes, hence, avoiding the inherently complex processing of the packet at the MAC layer and at the receiver \cite{simeone2008distributed}. One major approach for physical-layer synchronization is based on \acp{pco}, which use the reception times of the pulses transmitted by the other nodes to compute a correction signal applied to adjust the current node's \ac{vcc} \cite{simeone2008distributed,hong2005scalable, koskin2018generation}. In  classic \ac{pco}-based synchronization  \cite{simeone2008distributed}, 
  the correction signal is based on 
  the output of a \ac{pd} which computes
  the differences between the node's own time and the reception times of the pulses from the other nodes. 
  These differences are 
  weighted  according to the relative received pulse power w.r.t the sum of the  powers of the pulses received from the other nodes. 
  While with this intuitive weighting \ac{pco}-based synchronization is very attractive for wireless networks, 
  the resulting synchronization performance significantly degrade in network configurations in which there are large propagation delays and clock frequency differences, and generally, full clock synchronization (frequency and phase) is not attained by current \ac{pco}-based schemes, see, e.g., \cite{simeone2008distributed}.
  This motivates the design of a robust \ac{pco}-based time synchronization scheme, which is the focus of the current work.

  \textit{\textbf{Main Contributions}}: In this work we propose a \ac{pco}-based time synchronization scheme which is robust to propagation delays. To cope with the inherent challenge of mapping the  output of the \ac{pd} into a \ac{vcc} correction signal, we use a \ac{dnn}, building upon the ability of neural networks to learn complex mappings from data. To preserve the energy efficiency and distributed operation of \ac{pco}-based synchronization, we employ the model-based deep learning methodology \cite{shlezinger2020model,shlezinger2021model,shlezinger2022model}. Accordingly, our algorithm, coined  {\em \ac{dasa}}, augments the classic clock update rule of \cite[Eqn. (16)]{simeone2008distributed} via a dedicated \ac{dnn}. 
  In particular, we design \ac{dasa} based on the observation that conventional \ac{pco}-based synchronization is based on weighted averaging of the outputs of the \ac{pd}, which can be viewed as a form of self-attention mapping \cite{vaswani2017attention}. Thus, \ac{dasa} utilizes attention pooling, resulting in a trainable extension of the conventional algorithm. To train our model in a distributed fashion, we formulate a {\em decentralized loss measure} designed to facilitate rapid convergence, which can be computed at each node locally, resulting in a decentralized fast time synchronization algorithm. Our numerical results clearly demonstrate that the proposed \ac{dasa} yields rapid and accurate synchronization in various propagation environments, outperforming existing approaches in both convergence speed and performance. The proposed scheme is also very robust to values of the clock frequencies and to nodes' locations.
 
  
 \textit{\textbf{Organization}}: The rest of this work is organised as follows: Section \ref{sec:Clk_model} reviews the fundamental structure of \ac{pco}-based synchronization schemes.  Section \ref{sec:Preliminaries} presents the problem formulation, highlights the weaknesses of the classic weighting rule and states the objective of this work. Subsequently, Section \ref{sec:dnn_aided_synch} presents our proposed \ac{dasa}. Numerical examples and discussions are provided in Section~\ref{sec:results}. Lastly,   Section \ref{sec:conclusion} concludes this work.
 
 \textit{\textbf{Notations}}: In this paper, 
  deterministic column vectors are denoted with boldface lowercase letters, e.g., $\xvec$,  
  deterministic scalars are denoted via standard lowercase fonts,  e.g.,  $x$, and 
  sets are denoted with calligraphic letters, e.g., $\mX$. Uppercase Sans-Serif fonts represent matrices, e.g., $\mathsf{K}$, and the element at the $i$'th row and the $j$'th column of $\mathsf{K}$ is denoted with $\mathsf{K}_{i,j}$. The identity matrix is denoted by $\myMat{I}$. The sets of positive integers and of integers 
  are denoted by $\mathcal{N}$ and $\mZ$, 
  respectively. 
  Lastly, all logarithms are taken to base-2.

 \vspace{-0.2cm}
 
\section{Distributed Pulse-Coupled Time Synchronization for Wireless Networks}
\label{sec:Clk_model}
\vspace{-0.2cm}

\subsection{Network and Clock Models}
We study 
\ac{dt} clock synchronization for wireless networks, considering a network with $N$ nodes, indexed by $i\in\{1, 2, ..., N\}\triangleq\mI_N$. Each node has a clock oscillator with its own inherent period, denoted by $T_i$, $i\in\mI_N$. 
Generally, clock timing is often affected by an inherent random jitter, also referred to as phase noise. Let $V_i(k)$ denote the phase noise at node $i\in\mI_N$, at time index $k\in\mZ$.
Then, the corresponding clock time $\phi_i(k)$, can be expressed with respect to $k=0$ as 
	\begin{equation}
	    \phi_i(k)=\phi_i(0)+k\cdot T_i+V_i(k).
	\end{equation}
	
	\vspace{-0.2cm}
\noindent
In this work we  assume $V_i(k)=0$, $\forall k\in\mZ$ (see, e.g., \cite[Sec.V]{maggs2012consensus}, \cite{simeone2008distributed}) in order to focus on the fundamental factors affecting synchronization performance in wireless networks, which are the propagation delays and clock period differences.

\begin{figure}
		\centering
		\includegraphics[scale=0.70]{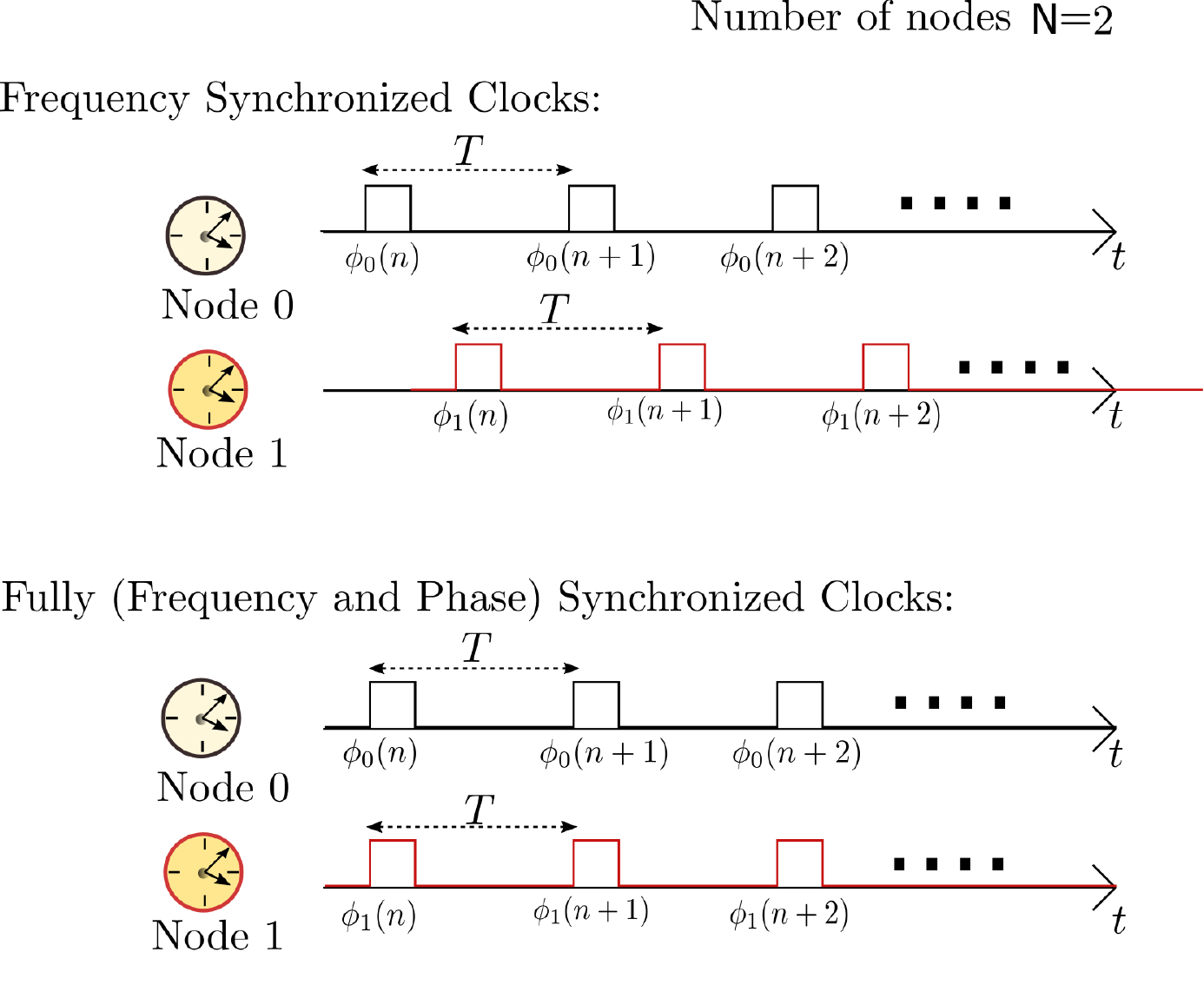}
		\caption{An illustration of clock frequency synchronization and of full clock synchronization, for a network which consists of two nodes.}
		\vspace{-0.6cm}
		\label{fig:types_of_synch}
	\end{figure}

In a wireless network, when the clock periods of the different nodes, $T_i$, $i\in\mI_N$, are different, then the nodes' transmissions may overlap in time and frequency (a situation which is referred to as ``collision"), resulting in loss of information. Moreover, even when the clock periods are identical, referred to as {\em clock frequency synchronization}, a time offset (also referred to as {\em phase offset}) between the clocks may exist, which again will result in collisions, as illustrated in Fig. \ref{fig:types_of_synch}.~Thus, to facilitate high speed communications, the nodes must synchronize both their {\em clock frequencies as well as their clock phases} to a common time base. This is referred to as {\em full clock synchronization}. To that aim, the nodes in the network exchange their current time stamps,~and, based on the exchanged time information, the nodes attempt to reach a consensus on  a common  clock.
	
 A wireless communications setup can be represented via a connectivity graph $\mathcal{X}$, consisting of a vertex set representing the nodes, and an edge set representing the links \cite[Ch. 1]{royle2001algebraic}. The edges between pairs of vertices (i.e., pair of nodes) are weighted by an adjacency matrix $\mathsf{A}$, whose $(i,j)$'th entry, $\left[\mathsf{A}\right]_{i,j}$, satisfies $\left[\mathsf{A}\right]_{i,j}\ge 0$, where $\left[\mathsf{A}\right]_{i,j}=0$ implies that there is no direct link between nodes $i$ and $j$.
	A connectivity graph has girth that is larger than one, hence the diagonal entries of $\myMat{A}$ are zero (i.e., $\left[\mathsf{A}\right]_{i,i}=0$). In the next subsections we  recall results on the convergence of \ac{pco}-based synchronization algorithms, obtained using the adjacency graph formulation, for specific cases discussed in the literature.
	
	Lastly, we note that in this work it is assumed that node locations are static and the propagation channels are time-invariant. The case of time-varying communications links has been studied in \cite{moreau2005stability, ren2005consensus, cao2012overview}, for which the adjacency matrix $\myMat{A}$ randomly evolves over time, and each node subsequently updates its coefficients following the information received from the other nodes.  In \cite{moreau2005stability},  necessary conditions for convergence were established by combining graph theory and system theory for bidirectional and for unidirectional links. It was concluded in \cite{moreau2005stability} that synchronization could fail, even for fully-connected networks.
	
	\subsection{Pulse-Coupled PLLs}
	\label{subsec:pulse-coupled-pll}
	
	As stated earlier, physical layer synchronization techniques operate by conveying the timing information of the nodes across to neighboring nodes via transmitted waveforms. Specifically, each node $i\in\mI_N$ broadcasts a sequence of synchronization signatures, which uniquely identifies the transmitting node, as in, e.g., \cite{alvarez2018distributed}, 
 where transmission times are determined at each transmitting node according to its own local clock. Each receiving node then processes the synchronization signatures received from all the other nodes, and updates its local clock using to a predetermined update rule. 
	
\begin{figure}
		\centering
		\includegraphics[width=\columnwidth]{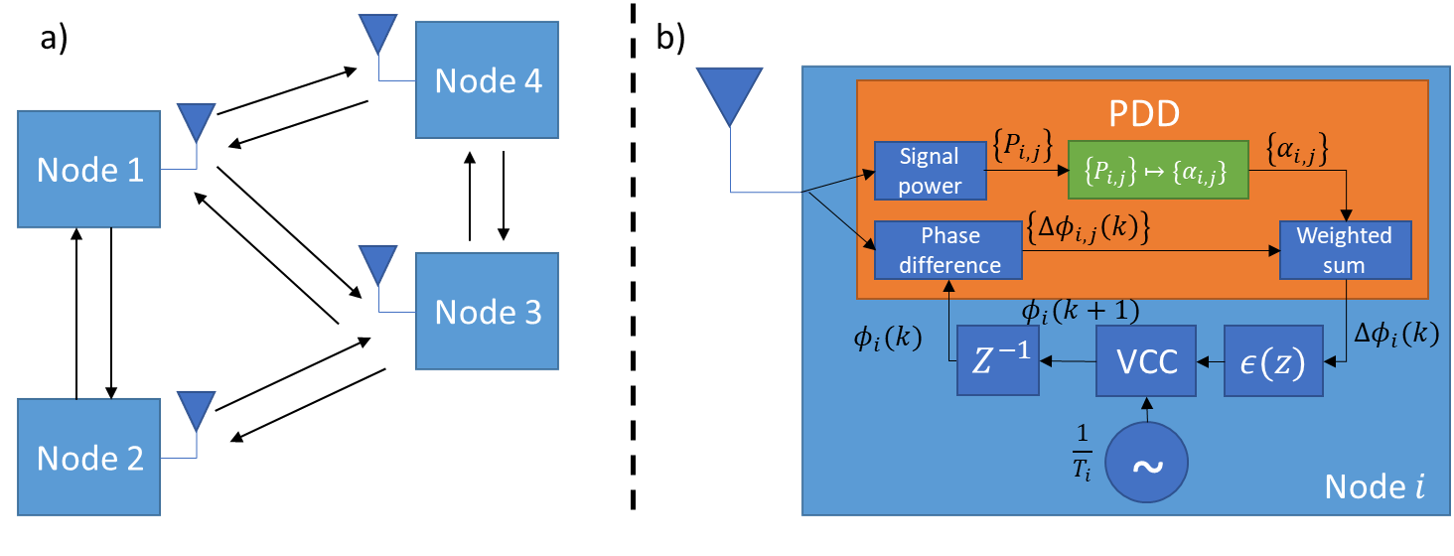}
 		\caption{Diagram showing (a) sample topological deployment of $N=4$ pulse-coupled nodes; (b) the structure of the synchronization mechanism based on the classical model \eqref{eqn:clock_analytical}, see \cite{simeone2008distributed}.} 
		\vspace{-0.6cm}
		\label{fig:DT_PLL}
	\end{figure}

	The distributed pulse-coupled \ac{pll} configuration is depicted in Fig. \ref{fig:DT_PLL}. It is assumed that the nodes operation is full-duplex, i.e., the nodes can transmit and receive at the same time. 
	At each node, the synchronization mechanism is based on a loop, which consists of a \ac{pdd}, a \ac{lti} filter with a transfer function $\varepsilon (z)$, and a \ac{vcc}. Each node is fed with the measured reception times of the pulses received from the different nodes, which are input to the \ac{pdd}. The \ac{pdd} calculates the difference between the time of each received pulse and the node's own clock, and weights this difference with an a-priori computed weighting factor, which is associated to the appropriate node based on its synchronization signature. The \ac{pdd} outputs the sum of the weighted differences to the loop filter $\varepsilon (z)$, which generates a correction signal for the \ac{vcc}. 
	Mathematically, the output of the \ac{pdd} at time index $k\in\mN$, at the $i$'th node, denoted by $\Delta \phi_i (k)$, can be expressed as:
	\begin{equation}
	\label{eqn: Time_difference}
	    \Delta \stampi (k)=\mathop{\sum}\limits_{j=1,j\neq i}^{N} \aij\cdot \left(t_{i,j}(k) -\stampi \left(k\right)\right),
	\end{equation}
	where $\aij\in [0,1]$, and $t_{i,j}(k)\triangleq \stampj\left(k\right) + q_{i,j}(k)$ is the reception time at node $i$ of the pulse transmitted by node $j$; which corresponds to the sum of the  transmission time, $\stampj\left(k\right)$, and the propagation delay $q_{i,j}(k)$ from node $j$ to node $i$. The \ac{pdd} output is then fed into a  loop filter $\varepsilon(z)$ whose output drives the \ac{vcc} that  re-calibrates the instantaneous time at the $i$'th node.

	\vspace{-0.3cm}

	\subsection{The Classic Pulse-Coupled PLL Configuration}
	\label{subsec:classicPCO}
	\vspace{-0.1cm}

	For the classic \ac{pco}-base \ac{pll} design of \cite{simeone2008distributed}, the $(i,j)$'th entry of the adjacency matrix $\mathsf{A}$ corresponds to the relative signal power of the pulse received at  node $i$ from  node $j$, with respect to the powers of all the other nodes received at node $i$: Denoting $\left[\mathsf{A}\right]_{i,j}=\aij$ and letting  $\pr$ denote the power of the pulse received at node $i$ from node $j$, then, in the classic algorithm of \cite{simeone2008distributed}, $\aij$ is computed as \cite{tong1995theoretical,simeone2008distributed,sourour1999mutual}:
	\begin{equation}
	\label{eqn:conn_weights}
	    \aij=\frac{\pr}{\mathop{\sum}\limits_{j=1,j\neq i}^{N} \pr}.
	\end{equation}
	From Eqn. \eqref{eqn:conn_weights} it follows that the value of $\aij$ depends on the distance between the nodes as well as on other factors which affect the received power levels, e.g., shadowing and fading. 
	When implementing a first-order \ac{pll}, then $\varepsilon (z)$ is set to $\varepsilon (z)=\varepsilon_0$, and letting $\Delta \phi_{i,j}(k)\triangleq \stampj\left(k\right) + q_{i,j}(k) -\stampi \left(k\right)$,  the update rule is (see \cite[Eqns. (16), (23)]{simeone2008distributed}):
	\begin{equation}
	\label{eqn:clock_analytical}
	    \stampi \left(k+1\right)=\stampi(k)+T_i+\varepsilon_0 \cdot \mathop{\sum}\limits_{j=1,j\neq i}^{N} \aij\cdot \Delta \phi_{i,j}(k).
	\end{equation}
    We  refer to the rule \eqref{eqn:clock_analytical} with weights \eqref{eqn:conn_weights} as the {\em classic algorithm} or the {\em analytic algorithm}.

	In this work, we investigate  distributed synchronization based on  \ac{dt} pulse-coupled \ac{pll}. With the adjacency matrix $\mathsf{A}$ defined above, the Laplacian matrix of the connectivity graph $\mathcal{X}$ is given as $\mathsf{L}=\mathsf{I}-\mathsf{A}$.
	It has been noted in \cite{simeone2008distributed} that
	for pulse-coupled first-order \ac{dt} \acp{pll}, 
	synchronization can be achieved if and only if $|\lambda_l(\myMat{L})|>0, \forall 2\le l \le N$, where $\lambda_l(\myMat{L})$ denotes the $l$'th eigenvalue of the matrix $\myMat{L}$, arranged in ascending order. 
	In general,  when using pulse-coupled \acp{pll}, synchronization across the nodes is attained when the connectivity graph is strongly connected; in other words, there should be a path connecting any node pair. The connection between each pair need not be direct, may also run via intermediate nodes, as long as all nodes in the network are able to exchange timing information among each other \cite{olfati2004consensus}. Hence, if there exists at least one node whose transmissions can be received at all the nodes in the network (directly or via intermediate nodes), then, clock frequency synchronization can be achieved.

The rule in \eqref{eqn:clock_analytical} was expressed as a time-invariant difference equation in \cite{simeone2008distributed}, for which the steady-state phase expressions for the nodes, in the limit as $k$ increases to infinity were derived. Specifically, for the case of no propagation delay and identical clock periods at all nodes, i.e., $q_{i,j}(k)=0$, $i,j\in\mI_N$, $k\in\mZ$, and $T_i=\Tnom$, $i\in\mI_N$, the rule \eqref{eqn:clock_analytical}  generally results in the network attaining full synchronization. On another hand, when there are propagation delays and/or different clock periods at the nodes, then typically, frequency synchronization to a common frequency is attained, but  full synchronization is not attained.
We consider in this paper the common and more practical scenario where there are propagation delays and different clock periods,  which generally results in asynchronous clocks at steady state. Accordingly, the objective of our algorithm is to attain full synchronization for this important scenario.

\vspace{-0.3cm}
\section{Problem Formulation}
\label{sec:Preliminaries}

    Consider a network with $N$ nodes, such that each node $i\in\mI_N$ has its own clock with an inherent period of $T_i$, where generally $T_i\ne T_j$ for $i\ne j$, and let $\phi_i(k)$ be the clock time at \ac{dt} time index $k\in\mN$. The nodes are located at geographically separate locations, where node $i$ is located at  coordinate $(x_i, y_i)$, and $\dij$ is the distance between nodes $i$ and $j$. Assuming line-of-sight propagation, a signal transmitted from node $i$ is received at node $j$ after $\qij=\frac{\dij}{c}$ seconds, where $c$ is the speed of light. We assume that the nodes are not aware of their relative  locations and of the  clock periods at  the other nodes. {\em The objective is to synchronize the clock times $\{\phi_i(k)\}_{i\in\mI_N}$ such that at steady-state, at each $k$, it holds that $\phi_i(k)=\phi_j(k)$, $\forall i\ne j$.} 
    
    	\begin{figure}[t]
		\centering
		\begin{subfigure}[t]{0.42\columnwidth}
		\centering
		\includegraphics[width=\textwidth]{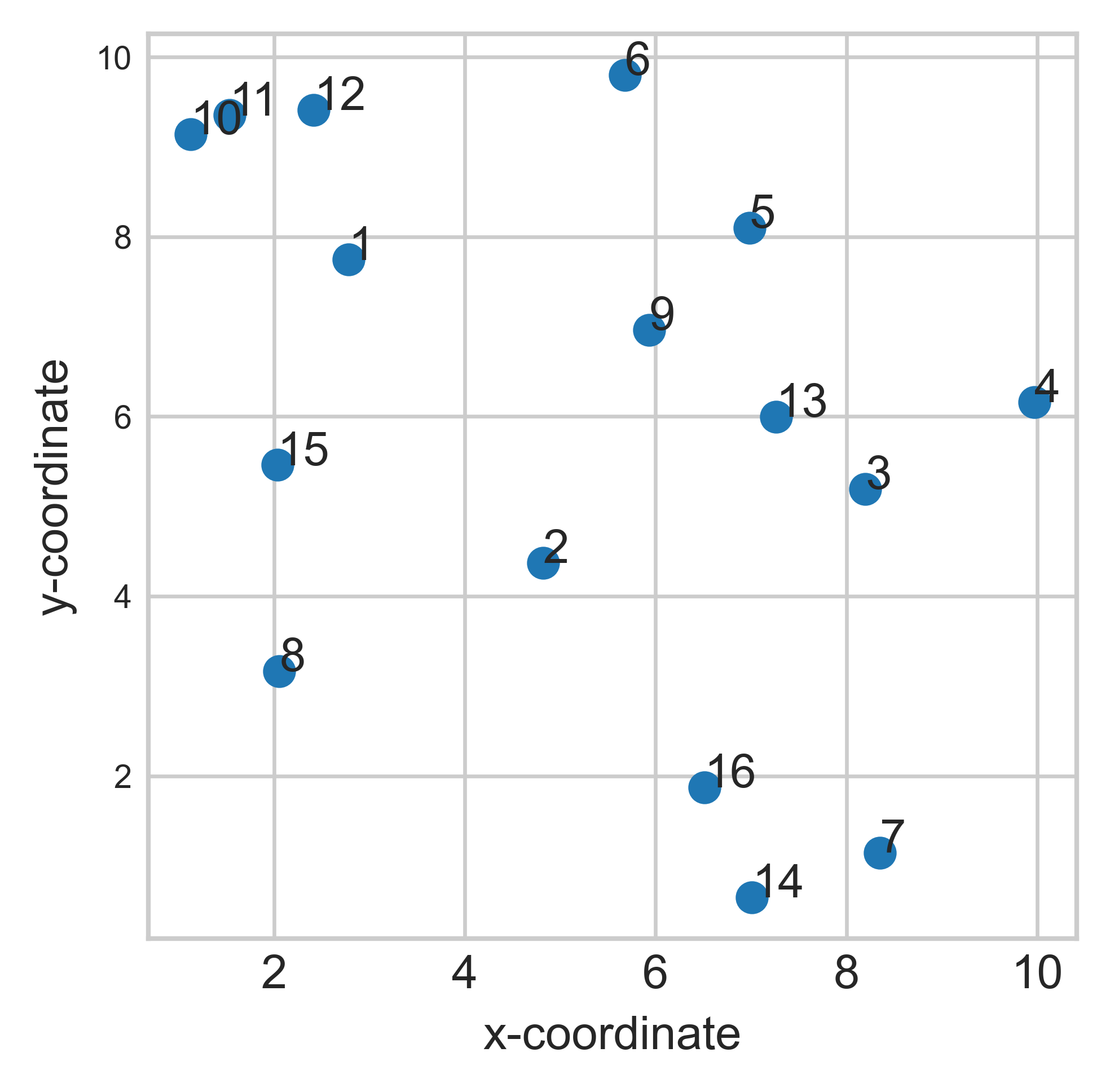}
		\caption{}
		\end{subfigure}
		\quad
		\begin{subfigure}[t]{0.42\columnwidth}
		\centering
		\includegraphics[width=\textwidth]{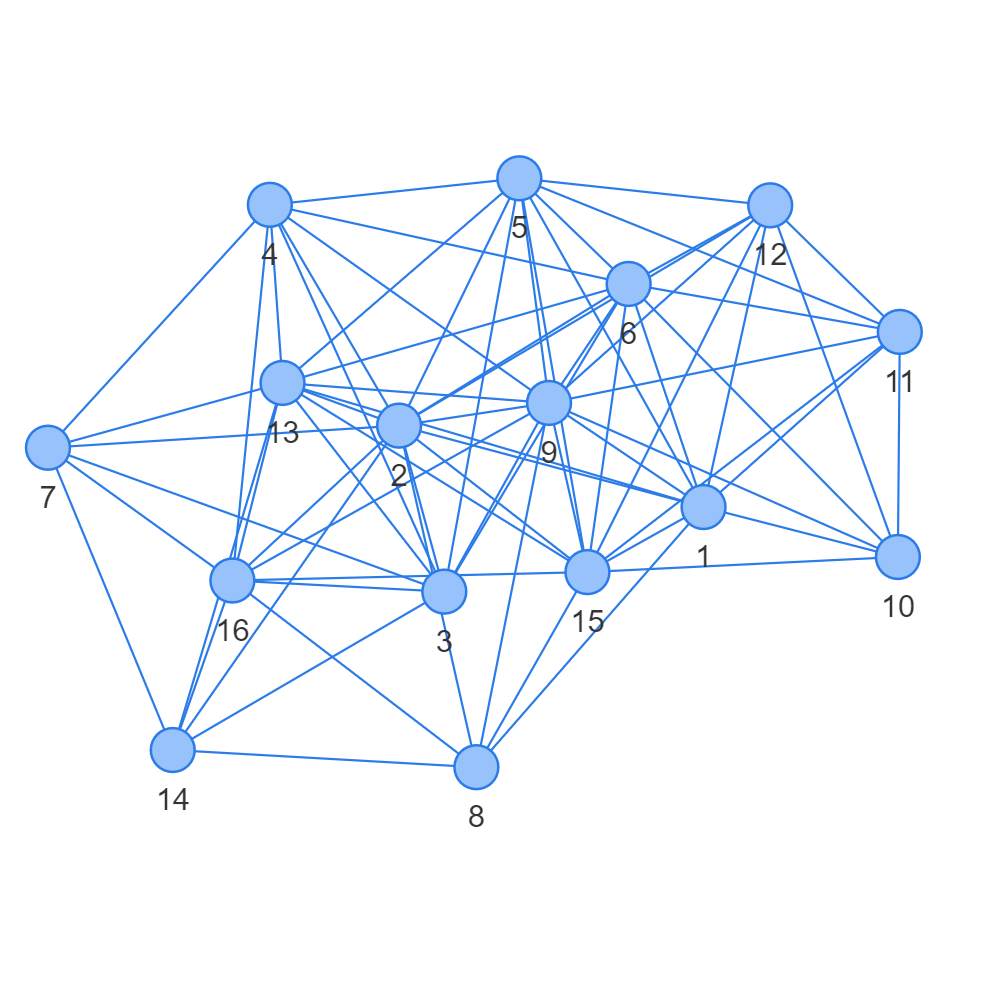}
		\caption{}
		\end{subfigure}
        \vspace{-0.3cm}
        \setlength{\belowcaptionskip}{-15pt}

		\caption{(a) Geographic locations of the nodes in the wireless network considered in Section \ref{sec:Preliminaries} and (b) Network graph showing the connected node pairs.}
		\label{fig:topology}
        \vspace{-0.2cm}
	\end{figure}
    
    To motivate the proposed approach we first illustrate the weakness of the analytic  update rule of \cite[Eqns. (16), (23)]{simeone2008distributed}), as discussed in Section \ref{subsec:pulse-coupled-pll}.
    This rule has been acceptable as a baseline rule in multiple works on network clock synchronization, e.g., \cite{tetreault2017use,pham2017stochastic,simeone2007distributed}, 
    hence, we use it  as a baseline for the performance of our proposed algorithm.
    As a motivating scenario, consider a wireless network with $N=16$ nodes located in a square area of dimensions $10 \;[\mbox{Km}] \times 10 \;[\mbox{Km}]$, with locations depicted in Fig. \ref{fig:topology}. 
    In this example, each node has a random clock time at startup \cite{simeone2008distributed}, taken uniformly over $[0,T_i]$, see, e.g., \cite{alvarez2018distributed,tetreault2017use, hulede2021distributed, amor2019joint}. 
    Each node transmits periodically at its corresponding clock times, and processes the pulse timing for its received pulses using the \ac{dt} \ac{pll} update rule  of \cite[Eqns. (16), (23)]{simeone2008distributed} to synchronize their clocks, see Eqn. \eqref{eqn:clock_analytical}.
    
    For the purpose of the numerical evaluation, we let the nominal 
    period of the clocks in the network, denoted  $\Tnom$, be  $\Tnom = \frac{1}{200}\;[\mbox{sec}]$. The period $T_i$ for the \ac{vcc} of node $i$ is obtained as by randomly generating clock periods with a maximum deviation of $100$ [ppm]: 
	\begin{equation}
	    T_i=\Tnom\cdot (1+B_0\cdot10^{-A}),
	\end{equation} 
	where $B_0$ is a uniformly distributed  random variable whose value is either $1$ or $-1$, and $A$ is uniformly selected from the interval $[4,6]$.  For time-invariant channels, the corresponding propagation delays    are given by $q_{i,j}(k)=q_{j,i}(k)=\frac{\dij}{c}$, $\forall i,j \in \mI_N$ and $k \in \mN$.

	For simplicity we assume identical transmit power of $P_t=33$ [dBm] at all nodes (different powers can be modeled using different topologies), and a two-ray wireless propagation model, in which
	the received signal consists of a direct \acl{los} component  and a single ground reflected wave component. Assuming isotropic antennas, the antenna gains are is $G_i=1$ at all directions, $\forall i\in\mI_N$. 
	For node heights of $1.5$ [m], it follows that the  received power at node $i$ from node $j$, denoted $P_{i,j}$, is given by the expression: \cite[Eqn. 2.1-8]{jakes1974microwave}:  
    \begin{equation}
	        P_{i,j}\approx \frac{P_t \cdot G_{j} G_{i} h_i^2 h_j^2}{\left(\dij\right)^4}=\frac{10}{\left(\dij\right)^4} = P_{j,i}.
	\end{equation}
	We assume receiver sensitivity of $-114$ [dBm], \cite{pala2015superregenerative}, \cite{naik2018lpwan}, \cite{scogna2017rfi} 
	and as a result, $48$ node pairs do not have direct reception.
	This is depicted in the graph in Fig. \ref{fig:topology}(b), in which nodes not connected by edges do not have a  direct link. The evaluated received power levels are then applied in calculation of the weights $\aij$ via Eqn. \eqref{eqn:conn_weights}. 
	
\begin{figure}
		\centering
		\begin{subfigure}[t]{0.43\columnwidth}
		\centering
		\includegraphics[width=\textwidth]{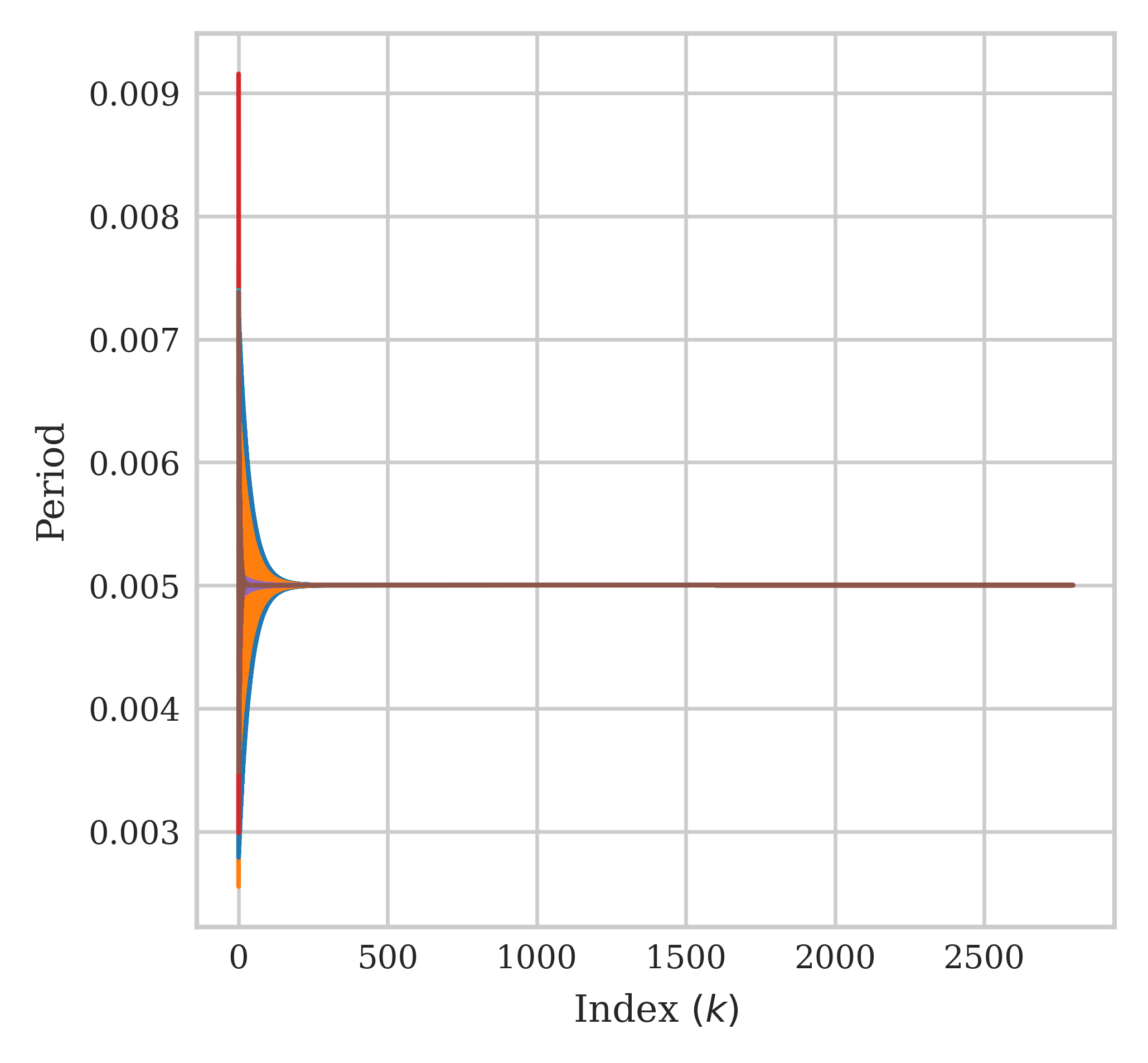}
    	\vspace{-0.6cm}
		\caption{}
		\label{fig:period_analytical_zeroorder}
		\end{subfigure}
		\quad
		\begin{subfigure}[t]{0.43\columnwidth}
		\centering
		\includegraphics[width=\textwidth]{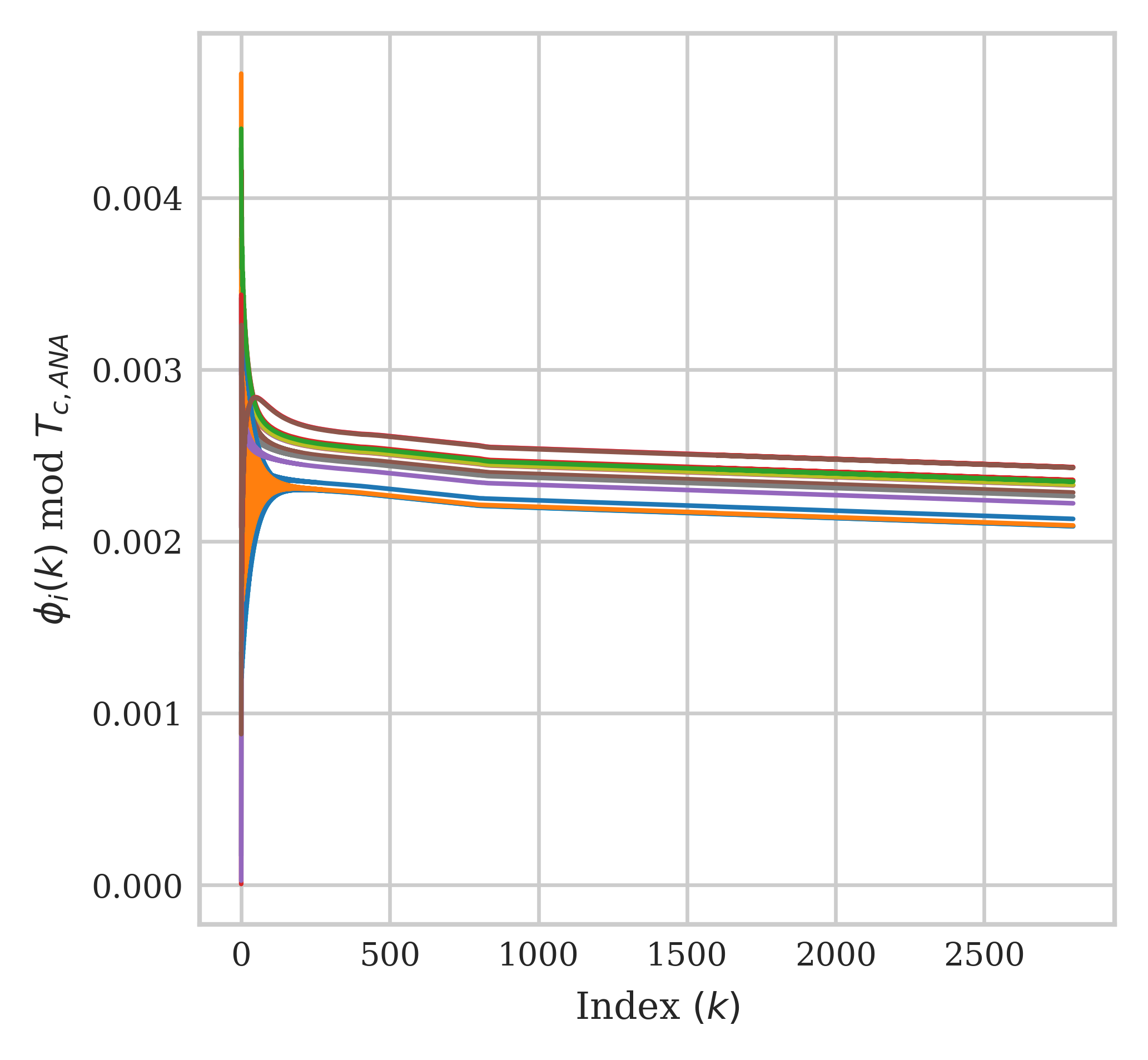}
		\vspace{-0.6cm}
		\caption{}
		\label{fig:phase_analytical_zeroorder}
		\end{subfigure}

\vspace{-0.3cm}
		\caption{Simulation results for the analytical model defined in Eqn. \eqref{eqn:clock_analytical} with a  first-order \ac{pll} $\varepsilon(z)=1$ (a) The evolution of the clock periods for all $16$ nodes; (b) Clock times $\stampi(k)$ modulo $T_{c,ANA}(2799)$, $i\in\mI_N$,  with fixed $\aij$'s computed via Eqn. \eqref{eqn:conn_weights}. 
		}
		\label{fig:zero_order}
        \vspace{-0.4cm}
	\end{figure}

\begin{figure}[t]
		\centering
		\begin{subfigure}[t]{0.43\columnwidth}
		\centering
		\includegraphics[width=\textwidth]{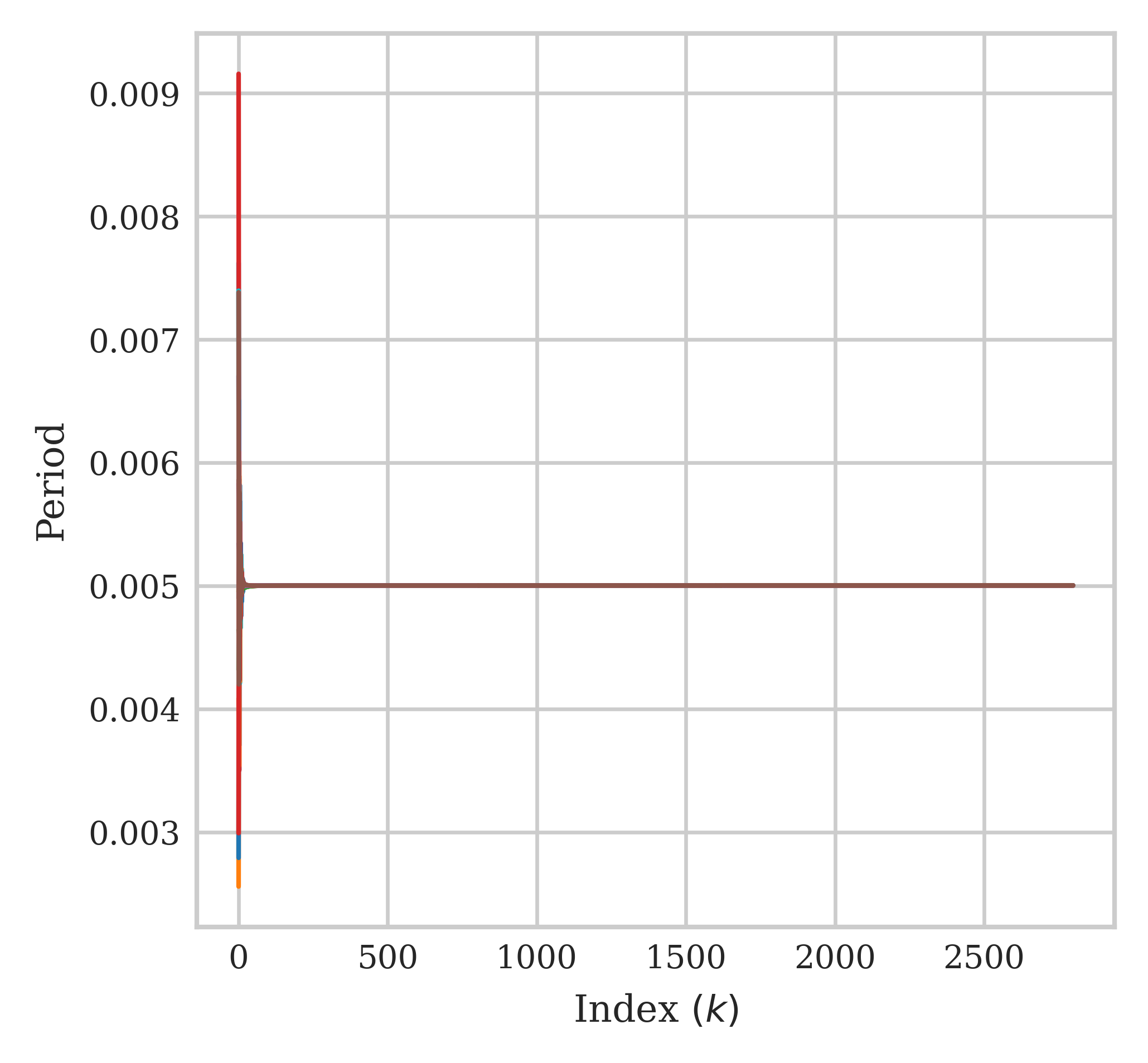}
		\vspace{-0.6cm}
		\caption{}
		\label{fig:period_analytical}
		\end{subfigure}
		\begin{subfigure}[t]{0.43\columnwidth}
		\centering
		\includegraphics[width=\textwidth]{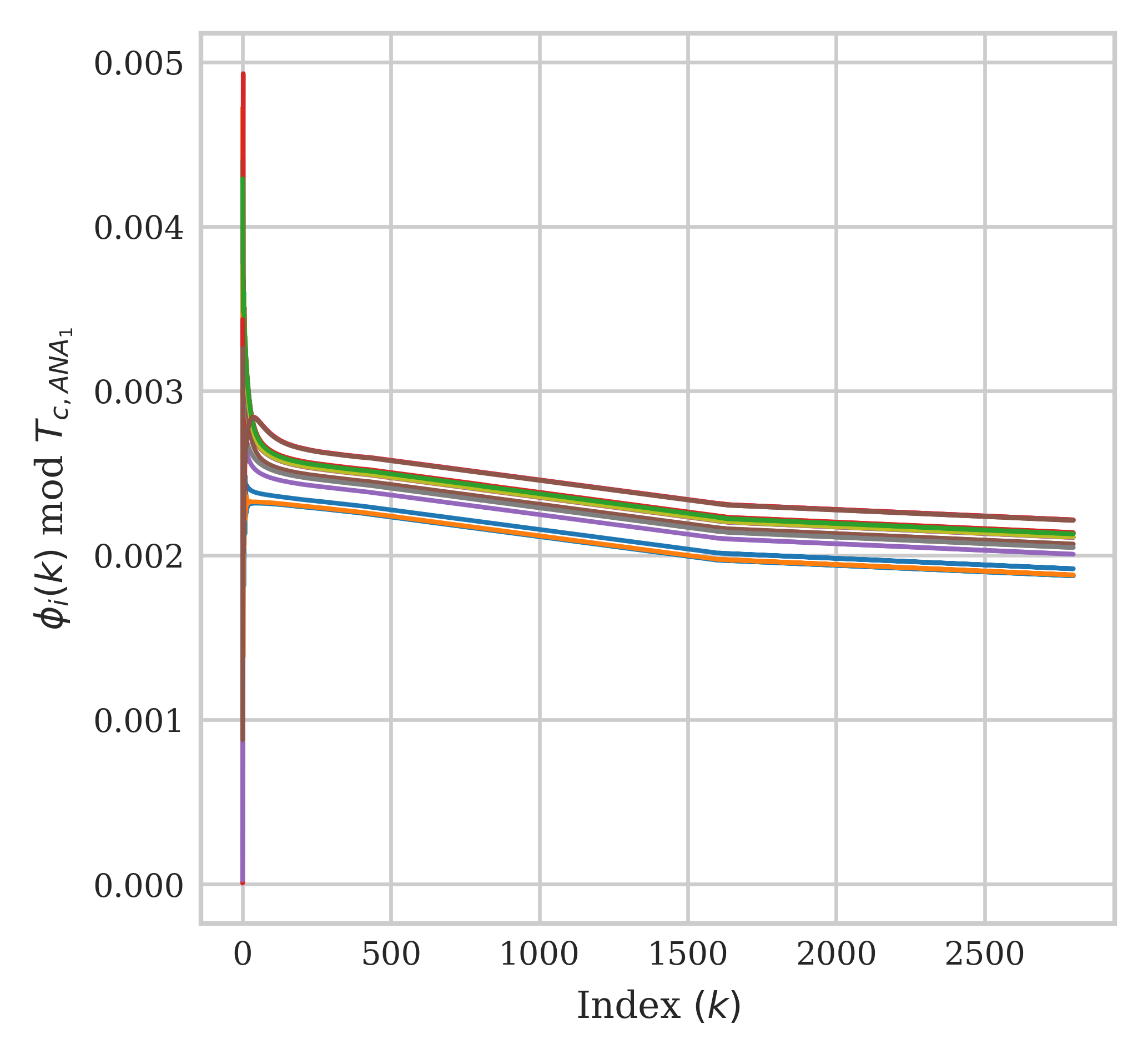}
		\vspace{-0.6cm}
		\caption{}
		\label{fig:phase_analytical_firstorder}
		\end{subfigure}
\vspace{-0.2cm}
\setlength{\belowcaptionskip}{-15pt}

		\caption{
		Simulation results for the analytical model defined in Eqn. \eqref{eqn:clock_analytical} with a  second order \ac{pll} $\varepsilon(z)=\frac{1}{1-0.3 z^{-1}}$ (a) The evolution of the clock periods for all $16$ nodes; (b) Clock times $\stampi(k)$ modulo $T_{c, ANA_1}(2799)$ $i\in\mI_N$,  with fixed $\aij$'s computed via Eqn. \eqref{eqn:conn_weights}}.

		\label{fig:first_order}
		\vspace{-0.6cm}
	\end{figure}

	In Figs. \ref{fig:zero_order} and \ref{fig:first_order} we evaluated the performance of the rule  \cite[Eqns. (16), (23)]{simeone2008distributed} for a first-order \ac{pll} with $\varepsilon(z)=1$ and for a second-order \ac{pll} with $\varepsilon(z)=\frac{1}{1-0.3 z^{-1}}$, respectively.  
	Figs. \ref{fig:period_analytical_zeroorder} and \ref{fig:period_analytical} depict the evolution of the {\em clock periods} for all the nodes, for the first-order \ac{pll} and the second-order \ac{pll}, respectively, as $k$ increases: The  period is observed to converge to a mean value at time $k=2799$, $T_{c,ANA}(2799)=0.00500208$, and  $T_{c,ANA_1}(2799)=0.00500274$, respectively,  which indeed demonstrates that frequency synchronization is achieved. The evolution of the modulus of the {\em clock  phases}  w.r.t the synchronized periods $T_{c,ANA}(2799)$ and $T_{c,ANA_1}(2799)$  when $\varepsilon(z)$ corresponds to a first-order \ac{pll} and to a second-order \ac{pll}, respectively, vs. $k$, is depicted in Figs. \ref{fig:phase_analytical_zeroorder} and \ref{fig:phase_analytical_firstorder}. Observe that indeed, for both \acp{pll}, different phases are achieved at steady state. 
	It follows that the algorithm \cite[Eqns. (16), (23)]{simeone2008distributed} achieves frequency synchronization but {\em not full synchronization}, resulting in collisions between the transmissions of the different nodes, which, in turn, reduce the throughput of the communications network.

    The examples above illustrate the motivation for our proposed solution: As the ad-hoc analytic expression of the weights does not lead to satisfactory synchronization performance when propagation delays and/or clock period differences exist, we propose to use a \ac{dnn}-aided mechanism to learn 
    the \ac{vcc} correction signals at the nodes, which lead to full network clock synchronization. In addition to  attaining the desired performance, attention is also given to {\em practicality} of implementation. Therefore, we require that the algorithm will operate in {\em a distributed manner}, such that each node  adapts its clock independently, processing only its own received signals. This is motivated by the fact that without independent processing, the network throughput  is further decreased due to the exchange of messages for facilitating coordinated processing. As the  update rule in Eqn. \eqref{eqn:clock_analytical} achieves partial synchronization and is plausible from an engineering perspective, our approach maintains the structure of the update rule, replacing only the weights $\aij$ with coefficients learned using a \ac{dnn}, where  the loss function used for training the \ac{dnn} coefficients is {\em calculated locally at each node}. This approach, referred to as {\em model-based learning}, is motivated by the fact that the coefficients computation \eqref{eqn:conn_weights} does not account for clock frequency differences between the nodes and does not fully account for propagation delays, thus it is reasonable to expect a trained \ac{dnn} to better account for these 
    factors.
    Model-based learning is also shown to carry the important advantage of robustness to the training set, as the learned functionality is restricted and does not entail end-to-end operation 
    \cite{shlezinger2021model,shlezinger2020model,shlezinger2022model}. In the next section we detail the proposed \ac{dnn}-based clock synchronization algorithm.


%
%
    \vspace{-0.2cm}
    \section{DNN-Aided Distributed Clock Synchronization Algorithm}
    \label{sec:dnn_aided_synch}
    \vspace{-0.2cm}
    
    In this section we present the proposed \ac{dasa}. We first describe the overall system in Section~\ref{subsec:Description}. Then, in Section~\ref{subsec:training} we detail how \ac{dasa} can be trained locally at each node in an unsupervised manner. We conclude with a discussion and a complexity analysis of the algorithm in Section~\ref{subsec:discussion}. 

	
	\vspace{-0.4cm}
    \subsection{Augmenting PLL-Based Synchronization with Deep Learning}
    \label{subsec:Description}
\begin{figure*}
		\centering
		\includegraphics[width=0.6\columnwidth]{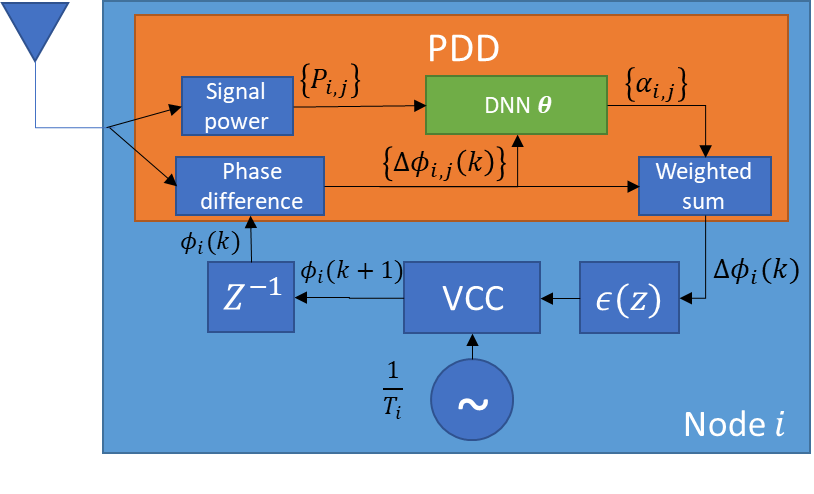}
		\caption{Block diagram of the proposed \ac{dasa}.}
		\vspace{-0.6cm}
		\setlength{\belowcaptionskip}{-20pt}
		\label{fig:dnn_pll_model}
	\end{figure*}

    Our proposed algorithm builds upon the classic \ac{pll}-based synchronization model, while augmenting its operation with a dedicated \ac{dnn}, in order to learn the appropriate weights for scenarios with non-negligible propagation delays and with clock frequency offsets. 
    Our design is inspired by attention mechanisms, which have demonstrated dramatic empirical success in natural language and in time sequence processing  \cite{vaswani2017attention}. Attention mechanisms are deep models that  learn to compute input-dependent weighting coefficients as a means of assigning increased significance to certain input sets and a lesser significance to others, which bears a clear similarity to the effect of the weighting coefficients $\{\aij\}$ on the \ac{pdd} output in \eqref{eqn:clock_analytical}. Consequently, 
    we replace the model-based weighting via $\{\aij\}$ in Eqn.~\eqref{eqn:conn_weights} with a \ac{dnn}-aided computation, where, again inspired by the success of self-attention mechanisms, we provide as inputs to the \ac{dnn} not only the power levels (as needed by 
    the classic algorithm, see Eqn. \eqref{eqn:conn_weights})  
    { but also the relative time differences $\{\Delta \phi_{i,j}(k)\}$.\black} The resulting algorithm, coined \ac{dasa}, is schematically depicted in Fig.~\ref{fig:dnn_pll_model}. \ac{dasa}  preserves the structure of the model-based \ac{dt} \ac{pll} synchronization scheme fo Eqn. \eqref{eqn:clock_analytical}, while harnessing the capabilities of the \ac{dnn} to learn abstract mappings from data, in order to replace the model-based weighting coefficients, which result in the classical algorithm being sensitive to propagation delays and clock frequency differences.

	 We denote the \ac{dnn} parameters  at node $i$ by $\myVec{\theta}_i(\cdot)$, and use $\psi_{\myVec{\theta}_i}$ to denote the resulting mapping. 
	 For a given value of $\myVec{\theta}_i$, the \ac{dnn} at node $i$ maps the $2(N-1)$ values $\big\{\Delta \phi_{i,j}(k), P_{i,j}(k)\big\}_{j=1, j\ne i}^N$ into the $N-1$ weighting coefficients $\{\aij\}$. Note that the coefficients $\{\aij\}$ also vary with~$k$.~The weighted sum of $\Delta \phi_{i,j}(k)$ is then input to a loop filter with a transfer function $\varepsilon(z)$, and the output of the loop filter drives the \ac{vcc}. The overall resulting time update rule can be expressed~as  
 	\begin{equation}
        \label{eqn:clock_analyticalDNN}
	    \stampi \left(k+1\right)=\stampi(k)+T_i+\varepsilon_0 \cdot \mathop{\sum}\limits_{j=1,j\neq i}^{N} \Big[\psi_{\myVec{\theta}_i}\Big(\Big\{\big(\Delta \phi_{i,j}(k), P_{i,j}(k)\big)\Big\}_{j=1, j\ne i}^N\Big) \Big]_j \cdot \Delta \phi_{i,j}(k),
	    	 \vspace{-0.05cm}
	\end{equation}
	 where $\big[\psi_{\myVec{\theta}_i}(\cdot)\big]_j$ denotes the output of the \ac{dnn} used for weighting the time difference between node $i$ and node $j$. 
	
	\subsubsection{Accounting for the Reception Threshold in DNN Structure}
	The fact that each receiver has a receive threshold below which it is not able to detect the existence of a signal, has to be accounted for in the design of the \ac{dnn}. Moreover, as the geographical locations of the nodes are unknown at the other nodes, {\em the effect of the detection threshold has to be handled without a-priori knowledge at the receiving nodes as to which are the nodes whose signal cannot be detected at each receiver}. Accordingly, it is not possible to a-priori set  the number of inputs at each \ac{dnn} to match the number of nodes received above the detection threshold. We thus set the number of \ac{dnn} inputs to $2(N-1)$ at all nodes. Then, whenever a transmitted pulse is reaches a given receiver below the detection threshold, we set both  corresponding input values of {\em receive power and phase difference} to $0$. 
	This can be implemented, e.g., by noting that signatures of certain users were not detected during a clock update cycle.
	As the \ac{dnn} outputs $N-1$ weights, then also in the calculation of the correction signal, the output weights corresponding to the timing of signals received below the detection threshold are set to zero.
	For example, if the pulse transmitted at time $k$ from node $j$ is not detected at node $i$ during the $k$'th clock update cycle, then we set \ac{dnn} inputs $P_{i,j}(k)=0$ and  $t_{i,j}(k)-\phi_i(k)=0$, and the \ac{dnn} output  $\Big[\psi_{\myVec{\theta}_i}\Big(\Big\{\big(\Delta \phi_{i,j}(k), P_{i,j}\big)\Big\}_{j=1, j\ne i}^N\Big) \Big]_j$ is multiplied by zero in the calculation of the update. 
	
While we draw inspiration from attention mechanisms, we note that our proposed \ac{dnn} is implemented as a \ac{mlp}, instead of using more sophisticated trainable attention mechanisms (e.g., multi-head attention \cite{vaswani2017attention}). This follows from the fact that the network size $N$ is assumed to be fixed, and thus there is no need to cope with inputs of varying lengths, as is the case in  multi-head attention models. This facilitates utilizing \acp{dnn} which can learn to exploit arbitrary dependencies between the inputs, while having a relatively low computational complexity and  being simple to train. 
The output of the \ac{mlp} is guaranteed to constitute weighted averaging coefficients by applying a softmax output layer with $N-1$ outputs.

			\vspace{-0.25cm}

	\subsection{Training Procedure}
	\label{subsec:training}
			\vspace{-0.1cm}

	\ac{dasa} is designed to support online training, making it robust and capable of coping with different environments and topologies. This is achieved via {\em unsupervised local training}, which can be carried out at each device locally without requiring access to some ground-truth clock values. We next elaborate on the training procedure, and subsequently discuss how one can acquire data for local training online. 
	
	\subsubsection{Unsupervised Local Training}
	Since training is done in an unsupervised manner, the training set at each user is  a sequence of $N_T$ \ac{dnn} inputs set, such each input set contains $N-1$  pairs of the receive time and received power level for the pulses received at a node from the other $N-1$ nodes. Accordingly, the training data set for the $i$'th node is given by
	\begin{equation}
	    \label{eqn:dataset}
	    \mySet{D}_i = \left\{ \left\{\big(t_{i,j}(k), P_{i,j}(k)\big) \right\}_{j=1, j\neq i}^{N} \right\}_{k=1}^{N_T}.
	\end{equation}
    
    The data set in \eqref{eqn:dataset} {\em does not contain any ground-truth clock value}. Nonetheless, it can still be used for training the algorithm to minimize the {\em relative time differences}, i.e., the differences between  each $t_{i,j}(k+1)$ and the clock time $\phi_i(k+1)$  produced by the \ac{dnn}-aided system after processing $\left\{\big(t_{i,j}(k), P_{i,k}(k)\big) \right\}_{j=1, j\neq i}^N$. 
    Since we are interested in achieving fast convergence, then offsets at earlier time instances are more tolerable compared with those obtained at later values of $k$. Accordingly, we weight the relative time differences in the computation of the loss function by a monotonically increasing function of $k$. Following \cite{samuel2019learning}, we use a logarithmic growth for weighting the $\ell_2$ loss. Consequently, the resulting loss function is given by
    \begin{equation}
        \label{eqn:loss1}
        \mySet{L}_{\mySet{D}_i}(\myVec{\theta}_i) = \sum_{k=1}^{N_T} \log(k+1) \bigg( \sum_{j =1, j \neq i}^{N} \left(\phi_i(k+1) - t_{i,j}(k+1)\right)^2\bigg), 
    \end{equation}
    with $\phi_i(k+1)$ computed recursively from $\phi_i(k)$ based on $\mySet{D}_i$ and $\myVec{\theta}_i$ via \eqref{eqn:clock_analyticalDNN}, i.e., 
    \begin{equation}
	\label{eqn:clock_analyticalDNN2}
	    \stampi \left(k+1\right)=\stampi(k)+T_i+\varepsilon_0 \cdot \mathop{\sum}\limits_{j=1,j\neq i}^{N} \Big[\psi_{\myVec{\theta}_i}\Big(\Big\{\big(t_{i,j}(k)-\stampi(k),  P_{i,j}(k)\big)\Big\}_{j=1, j\ne i}^N\Big) \Big]_j \cdot (\stampi(k) - t_{i,j}(k)).
    \end{equation}
    The fact that loss in \eqref{eqn:loss1} is a quadratic function of $\phi_i(k+1)$, which in turn is a linear recursive function of the \ac{dnn} output via \eqref{eqn:clock_analyticalDNN2}, indicates that one can compute the gradient of the loss with respect to the weights via backpropagation through time \cite{sutskever2013training}.
    
    We also note that the loss \eqref{eqn:loss1} can be computed in an unsupervised manner by each node locally, i.e., it relies only on the local \ac{dnn} of node $i$. Hence, it facilitates unsupervised local training via conventional first-order based optimizers. The overall training algorithm when utilizing the conventional gradient descent optimizer is summarized as Algorithm~\ref{alg:UnsupLocTraining_alg}.

\RestyleAlgo{ruled}
    \begin{algorithm}
    \caption{Unsupervised Local Training at Node $i$
    }\label{alg:UnsupLocTraining_alg} 
    \KwData{Data set $\mySet{D}_i$, learning rate $\mu$, initial weights $\myVec{\theta}_i$,  period $T_i$, number of epochs $E$}
    \For{${\rm epoch}=1$ to $E$}{ 
    \For{$k=1$ to $N_T$}{
    \textbf{Forward pass}   $\left\{\big(t_{i,j}(k), P_{i,j}(k)\big) \right\}_{j=1, j\neq i}^{N}  \in \mySet{D}_i$ to obtain $\phi_i(k+1)$ using Eqn. \eqref{eqn:clock_analyticalDNN2}.
    }
    \textbf{Compute loss} $\mySet{L}_{\mySet{D}_i}(\myVec{\theta}_i)$ via Eqn. \eqref{eqn:loss1}\;
    \textbf{Compute gradient} $\nabla_{\myVec{\theta}_i}\mySet{L}_{\mySet{D}_i}(\myVec{\theta}_i)$ using backpropagation through time\;
    \textbf{Update weights} via $\myVec{\theta}_i \leftarrow \myVec{\theta}_i - \mu \cdot \nabla_{\myVec{\theta}_i}\mySet{L}_{\mySet{D}_i}(\myVec{\theta}_i)$.
    }
    \end{algorithm}

	\subsubsection{Data Acquisition and DNN Training}
	The conventional practice in deep learning is to train  \acp{dnn} offline, using pre-acquired data for training, and then use the trained weights for inference on the deployed devices. For \ac{dnn}-aided clock-synchronization in wireless networks, training should  consider the specific propagation delays of the deployed network
	and clock frequency differences, as it is known from \cite{simeone2008distributed} that in the absence of these two factors the algorithm \eqref{eqn:conn_weights},  \eqref{eqn:clock_analytical} achieves full synchronization. While one may acquire data from measurements corresponding to the expected deployment and use it to train offline, a practically likely scenario is that the nodes will be required to train after deployment. 
	
	The training procedure in Algorithm~\ref{alg:UnsupLocTraining_alg} is particularly tailored to support on-device training, as it does not require ground-truth clock values and can be carried out locally. However, it still relies on providing each node with the training data set $\mySet{D}_i$ in \eqref{eqn:dataset}. Nonetheless, such data can be acquired by simply having each node submit a sequence of $N_T$ pulses, which the remaining nodes utilize to form their corresponding data sets. In particular, once the network is deployed and powered up, each device transmits $N_T$ pulses, and uses its received measurements to form its local data set $ \mySet{D}_i$. This step is carried out when the nodes are not synchronized.
	It is emphasized that during the data acquisition, the nodes do not update their \ac{dnn} coefficients, thus the parameters $\myVec{\theta}_i$ at node $i$ during this step are fixed to those obtained at the initialization. 
	Then, in the local unsupervised training step, each node trains its local \ac{dnn} via Algorithm~\ref{alg:UnsupLocTraining_alg}, using the acquired data $\mySet{D}_i$. 
	This results in the nodes having both synchronized clocks at time instance $N_T$, as well as trained weights $\{\myVec{\theta}_i\}$. 
	The trained model coefficients are then applied to compute the $\aij$'s, instead of the $\aij$'s of Eqn. \eqref{eqn:conn_weights}, {\em without requiring additional samples to be acquired and without re-training,} i.e., operating in a one-shot manner without inducing notable overheard. This local training method thus differs from \ac{drl} approaches, where training is carried out by repeated interaction, which in our case can be viewed as multiple iterations of data acquisition and local training.

	\vspace{-0.3cm}
	\subsection{Discussion}
	\label{subsec:discussion}
\begin{figure}
		\centering
		\includegraphics[scale=0.75]{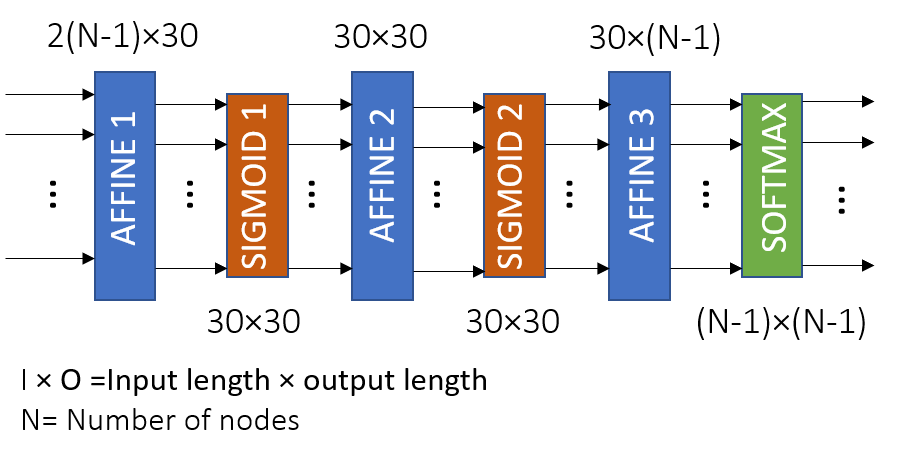}
		\vspace{-0.3cm}
		\caption{Block diagram of the \ac{dnn} utilized by the proposed \ac{dasa} in the experimental study reported in Section~\ref{sec:results}.}
		\vspace{-0.8cm}
		\label{fig:dnn_layers}
	\end{figure}

	The proposed \ac{dasa} learns from data the optimal implementation of \ac{pll}-based synchronization. We augment the operation of the model-based synchronization method of \cite{simeone2008distributed} to overcome its inherent sensitivity to non-negligible propagation delays and to clock frequency differences, harnessing the ability of \acp{dnn} to learn complex mappings from data. While we are inspired by attention mechanisms, which typically employ complex highly-parameterized models, \ac{dasa} supports the usage of compact, low-complexity \acp{dnn}, that are likely to be applicable on hardware-limited wireless devices. For instance, in our numerical study reported in Section~\ref{sec:results}, we utilized the simple three-layer \ac{mlp} illustrated in Fig.~\ref{fig:dnn_layers}, which is comprised of solely $(3(N-1)+30)\cdot 30$ weights and $2\cdot 30 + N-1$ biases; for a network with $N=16$ nodes that boils down to merely $2.5\cdot 10^3$ parameters -- much fewer than the orders of parameters of \acp{dnn} used in traditional deep learning domains such as computer vision and natural language processing.
	The application of a low-complexity \ac{dnn} augmented into an established algorithm also yields  a relatively low-complexity operation during inference, i.e., using the trained \ac{dnn} for maintaining synchronization. For instance, the instance of the aforementioned implementation with  $2.5\cdot 10^3$ parameters
	corresponds to fewer than  $2.5\cdot 10^3$  products on inference --  a computational burden which is likely to be feasible on real-time on modern micro-controllers, and can support parallelization implemented by dedicated \ac{dnn} hardware accelerators \cite{deng2020model}. 
	
	Our proposed training scheme bears some similarity to techniques utilized in multi-agent \ac{drl}, which acquire data by repeated interactions between distributed agents and the environment. However, our proposed method avoids the repeated interactions utilized in \ac{drl}, which in the context of clock synchronization would imply a multitude of exchanges of $N_T$ pulses among the nodes, leading to a decrease in network throughput. In particular, our proposed method enables nodes to learn the optimal synchronization parameters from a single sequence of transmitted pulses, and the trained \acp{dnn} can be subsequently employed at the respective nodes to maintain full clock (frequency and phase) synchronization between the nodes in the network. Nonetheless, in a dynamic network scenarios with highly mobile nodes, it is likely that the nodes may need to retrain their local models whenever the topology changes considerably from the one used during its training. We expect training schemes designed for facilitating online re-training in rapidly time-varying environments by, e.g., leveraging data from past topologies to predict future variations as in \cite{raviv2021meta,raviv2022online}; however, we leave these extensions of \ac{dasa} for future work. 
	
	
	

\vspace{-0.3cm}
\section{Performance Evaluation}
\label{sec:results}
\vspace{-0.2cm}

In this section we report an extensive simulation study to evaluate 
the performance of the proposed algorithm\footnote{The complete source code is available at \url{https://github.com/EmekaGdswill/Distributed-DNN-aided-time-synchronization.git}}, schematically described in Figs. \ref{fig:dnn_pll_model} and \ref{fig:dnn_layers}.  
 To facilitate a fair comparison between the \ac{dasa} and the classic algorithm \eqref{eqn:clock_analytical}, the parameters (i.e., $\pt, q_{i,j}, \pr, \left\{\stampi(0)\right\}_{i=1}^{N}$) {\em are identical} for the tests of  both the analytic algorithm \eqref{eqn:clock_analytical} and \ac{dasa}.  
We recall that it was shown in Section \ref{sec:Preliminaries} that the analytic algorithm fails to achieve full synchronization for the considered scenario.
\ac{dasa} consists of three steps: 1) Data acquisition step; 2) Training step; and 3)  Free-run or  testing step. At the free-run step, 
the nodes use their trained \acp{dnn} to update their clocks via the update rule \eqref{eqn:clock_analyticalDNN2} with their measured $\pr(k)$'s and $t_{i,j}(k)$'s.
	

	At startup, corresponding to clock index $k=0$, each node $i$, $i \in \mI_N$, obtains its initial clock time $\stampi(0)$,   generated randomly and uniformly over $[0,T_i]$ (see Section \ref{sec:Preliminaries}), and the \ac{dnn} parameters $\myVec{\theta}_i$ are initialized randomly and uniformly according to the PyTorch default setting.  
	Subsequently, the data acquisition step is applied at all nodes simultaneously. In this step, the nodes compute their clock times for pulse transmissions according to the update rule \eqref{eqn:clock_analyticalDNN2}, where the outputs of the local \acp{dnn} at the nodes are computed with the corresponding {\em   randomly initialized  parameters}, $\myVec{\theta}_i$, $i\in\mI_N$, {\em which are not updated} during the this step. 
	We set the duration of the
	data acquisition interval to $N_T=10$ reception cycles. At the end of the data acquisition interval, each node $i$ has a training data set $\mySet{D}_i$. 
	Next, the training step in applied at each node individually, where node $i$ uses the data set $\mySet{D}_i$ to train its individual  \ac{dnn}, $\psi_{\myVec{\theta_i}}$,  according to  Algorithm \ref{alg:UnsupLocTraining_alg}.
	It is emphasized that the data acquisition and the training processes {\em are carried out simultaneously} at the individual nodes, as each node applies processing based only on {\em its received pulse timings and powers}. We apply training over $E$ epochs, determined such that the individual loss per node  $\mySet{L}_{\mySet{D}_i}(\myVec{\theta}_i)$, defined in Eqn. \eqref{eqn:loss1}, reaches the asymptotic value. 
	After learning the parameters for each \ac{dnn}, $\myVec{\theta}_i$, 
	is completed, each node $i$ then continues to update its clock using the rule \eqref{eqn:clock_analyticalDNN2}
	with weights $\aij$ computed by applying the trained \ac{dnn} to its input data. At time $k$, the $N-1$ \ac{dnn} outputs at node $i$ are computed by $\psi_{\myVec{\theta}_i}\Big(\Big\{\big(t_{i,j}(k)-\stampi(k), P_{i,j}(k)\big)\Big\}_{j=1, j\ne i}^N\Big)$. 
	In the evaluations, we apply the testing step for $2800$ time indexes. 
	
\begin{figure} 
		\centering
		\begin{subfigure}[t]{0.43\columnwidth}
		\centering
		\includegraphics[width=\textwidth]{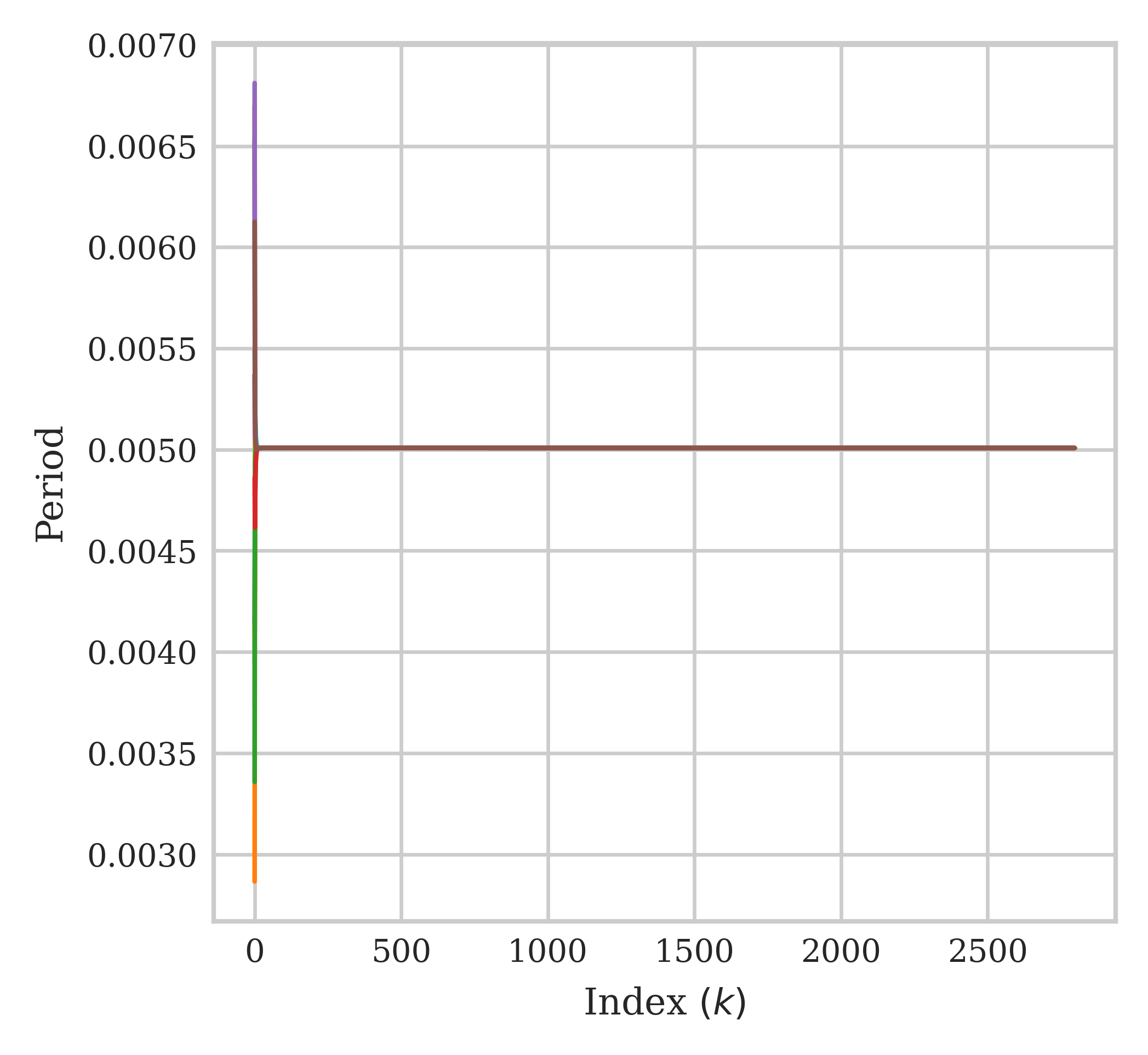}
		\vspace{-0.6cm}
		\caption{}
		\label{fig:Period_DNN_all}
		\end{subfigure}
		\quad
		\begin{subfigure}[t]{0.45\columnwidth}
		\centering
		\includegraphics[width=\textwidth]{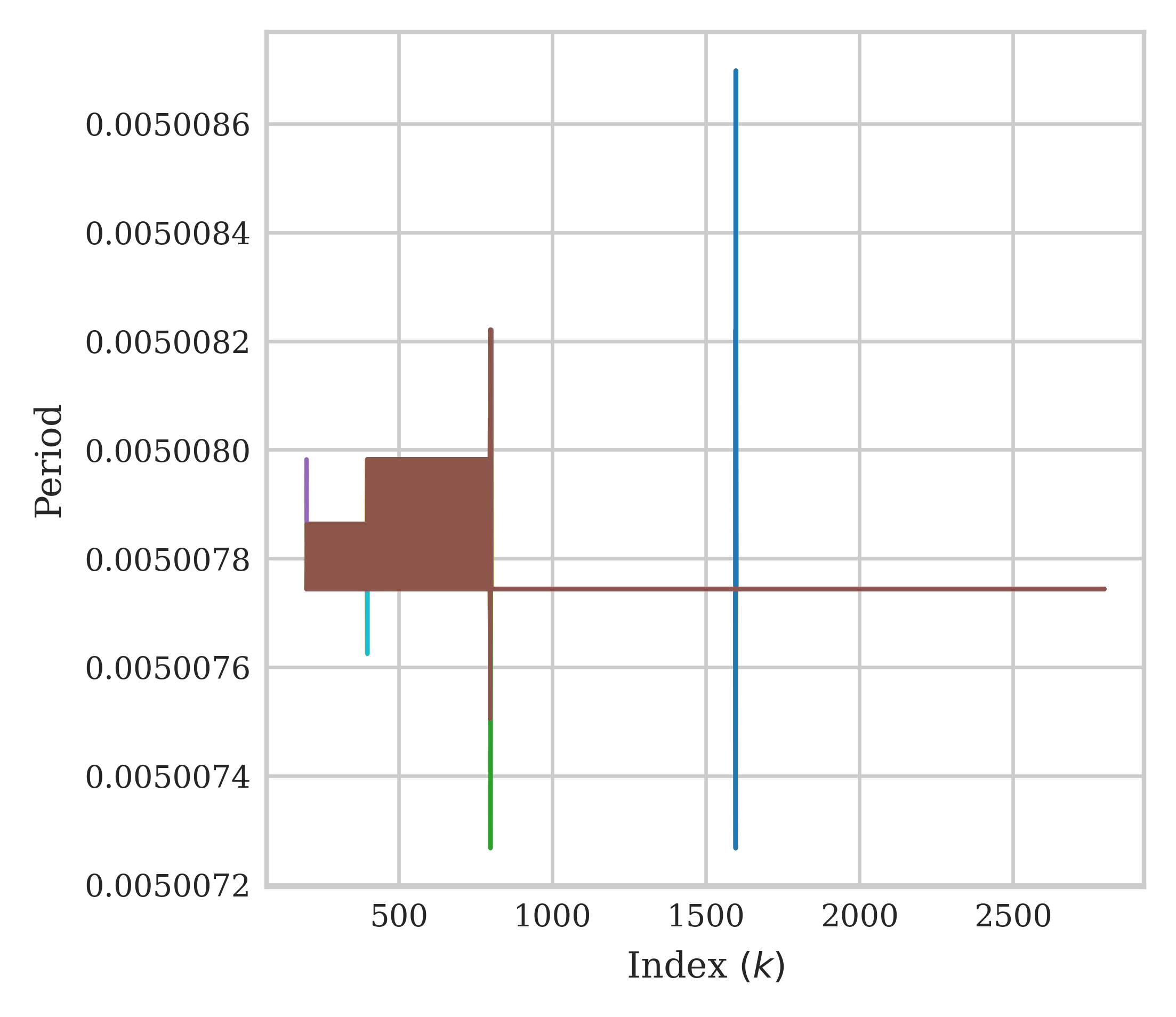}
		\vspace{-0.6cm}
		\caption{}
		\label{fig:Period_DNN_last}
		\end{subfigure}
		\vspace{-0.3cm}
		\caption{The evolution of the clock periods for all $16$ nodes using the proposed \ac{dasa}:  (a) The entire test period (time indices  0-2799); and (b) Zoom on the last 2600 time indices (time indices 200-2799).} 
		\label{fig:Period_DNN}
	\end{figure}

\begin{figure} 
	    \centering
        \begin{subfigure}[t]{0.43\columnwidth}
		\centering
		\includegraphics[width=\textwidth]{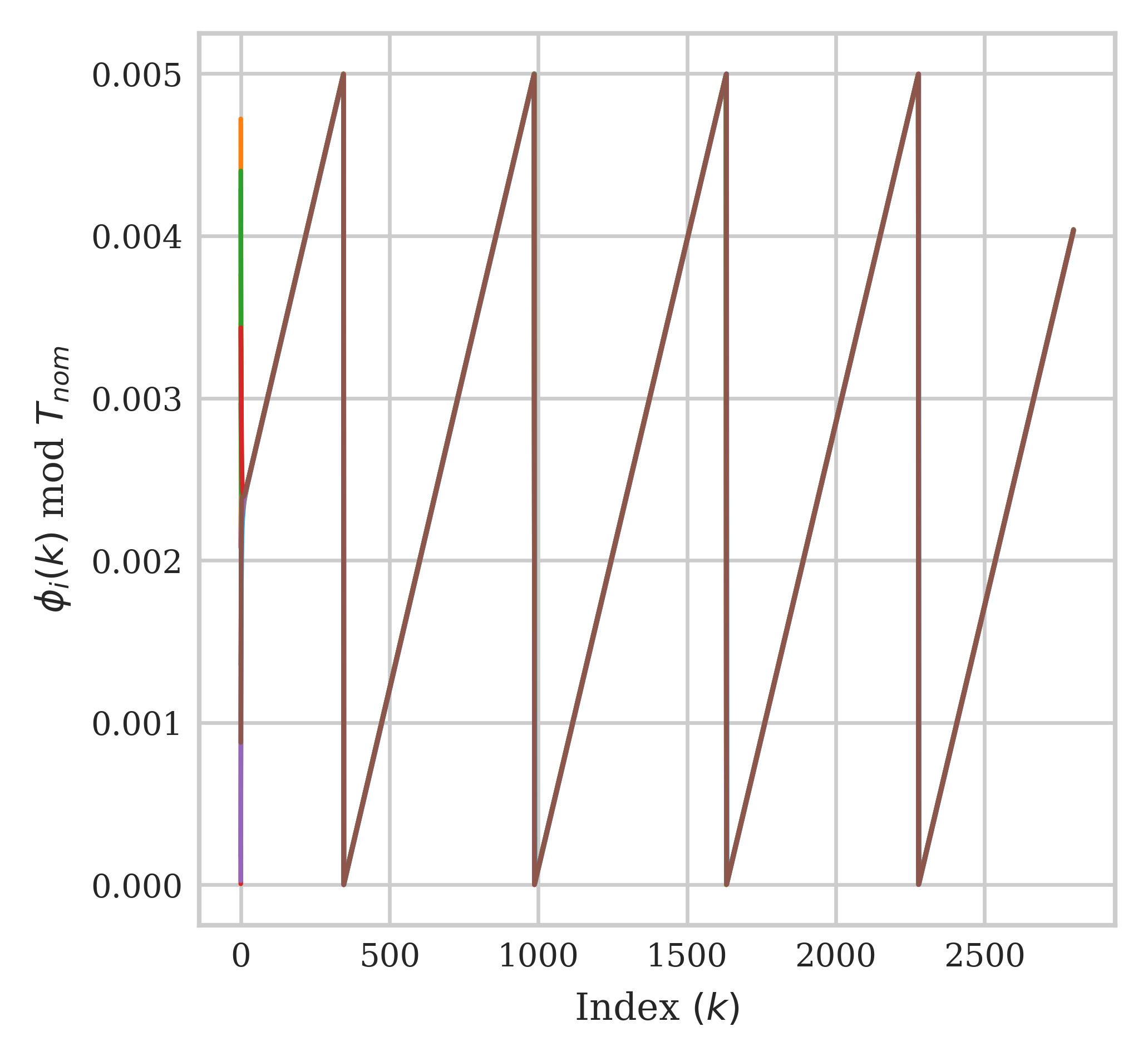}
		\vspace{-0.6cm}
		\caption{}
		\label{fig:phase_Neural_Tnom}
		\end{subfigure}	
		\quad
		\begin{subfigure}[t]{0.43\columnwidth}
		\centering
		\includegraphics[width=\textwidth]{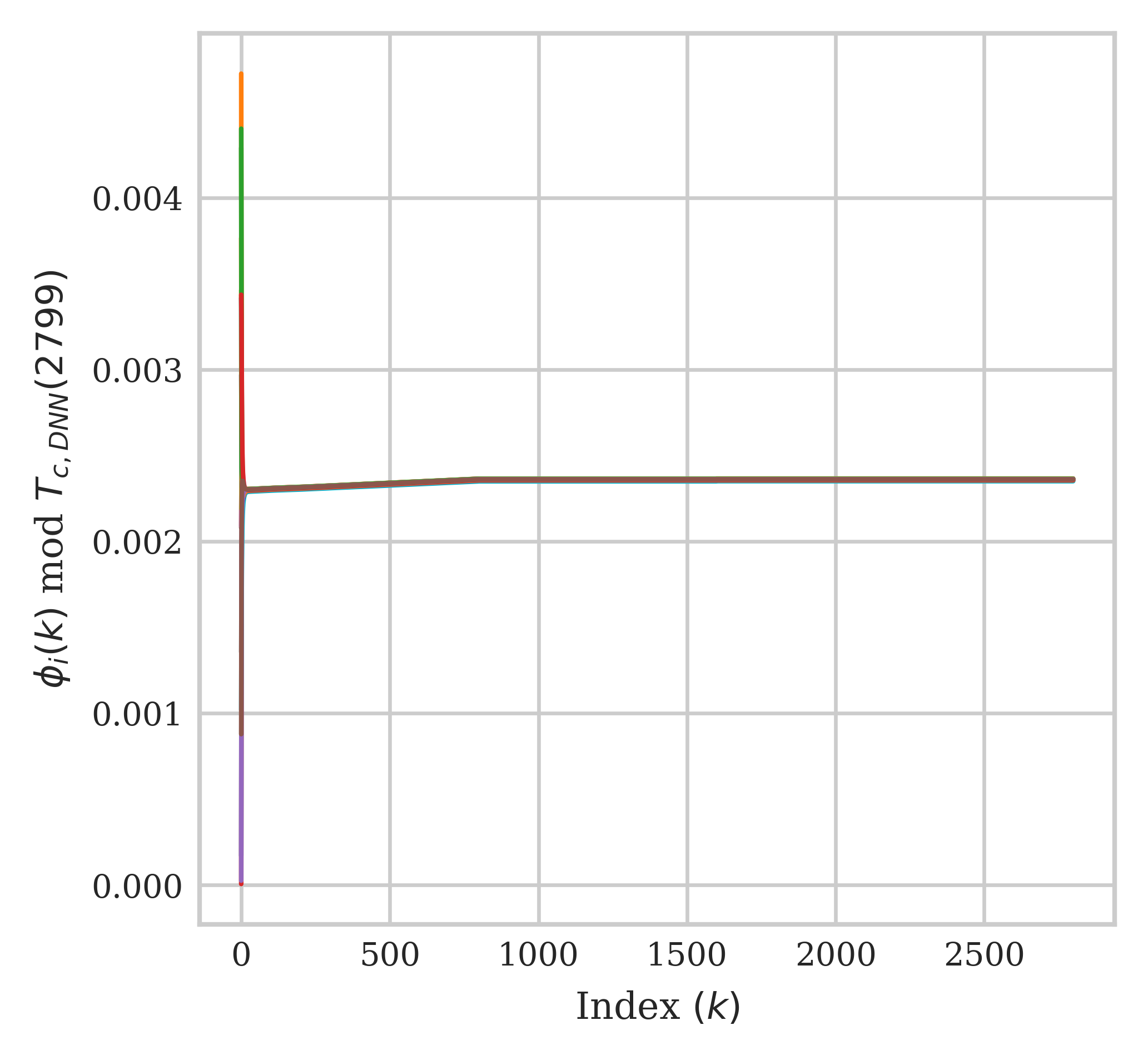}
		\vspace{-0.6cm}
		\caption{}
		\label{fig:phase_Neural_TDNN}
		\end{subfigure}
		\vspace{-0.3cm}
		\caption{ Modulus of clock phases $\stampi(k)$ for all $N=16$ clocks versus index $k$ using the proposed 
		\ac{dasa},
		  (a) w.r.t $\Tnom$  and (b) w.r.t $T_{c,DNN}(2799)$.}
		\label{fig:phase_Neural}
		\vspace*{-0.5cm}
	\end{figure}

	From the numerical evaluations we identified that setting $E=400$ epochs is sufficient for securing convergence. Recalling that each epoch corresponds to a single iteration, it is concluded that {\em convergence is relatively fast}. We first consider the behaviour of the clock period after training, 
	for the same network topology with $N=16$ nodes, considered in Section \ref{sec:Preliminaries} (see Fig.~\ref{fig:topology}), 
	depicted in Fig. \ref{fig:Period_DNN} for all $16$ nodes. From the time evolution in Fig. \ref{fig:Period_DNN_all} it is indeed observed that the nodes' period convergence is very quick. In  Fig. \ref{fig:Period_DNN_last} we focus in the last $2600$ clock indexes of the testing step: Observe that after convergence, there are still small jumps in the period, which are much smaller than the mean value of the converged period, i.e., $6$ orders of magnitude smaller, hence are considered negligible. It is also interesting to see that once one node experiences a jump in the period, then all other nodes follow.
	We obtain  that at the end of the testing step, the network attains a {\em mean synchronized period} of  $T_{c,DNN}(2799)=0.00500774$ (computed at the last testing index). Fig. \ref{fig:phase_Neural_Tnom} depicts the modulus of the clock phases w.r.t $\Tnom$ across all the nodes, and Fig. \ref{fig:phase_Neural_TDNN} depicts the modulus of the clock phases w.r.t $T_{c,DNN}$ across all the nodes. It is evident from the figure that the \ac{dnn}-aided network attains full synchronization w.r.t. $T_{c,DNN}$, which is different from $\Tnom$. 
	Comparing  Fig.  \ref{fig:phase_analytical_zeroorder} with Fig. \ref{fig:phase_Neural_TDNN} we conclude that the {\em proposed \ac{dasa} offers significantly better performance than the classical approach}.
	Moreover, the performance achieved using the trained \ac{dnn}
	is {\em  robust} to clock period differences and propagation delays. 


\begin{figure}[t]
		\centering
		\begin{subfigure}[t]{0.43\columnwidth}
		\centering
		\includegraphics[width=\textwidth]{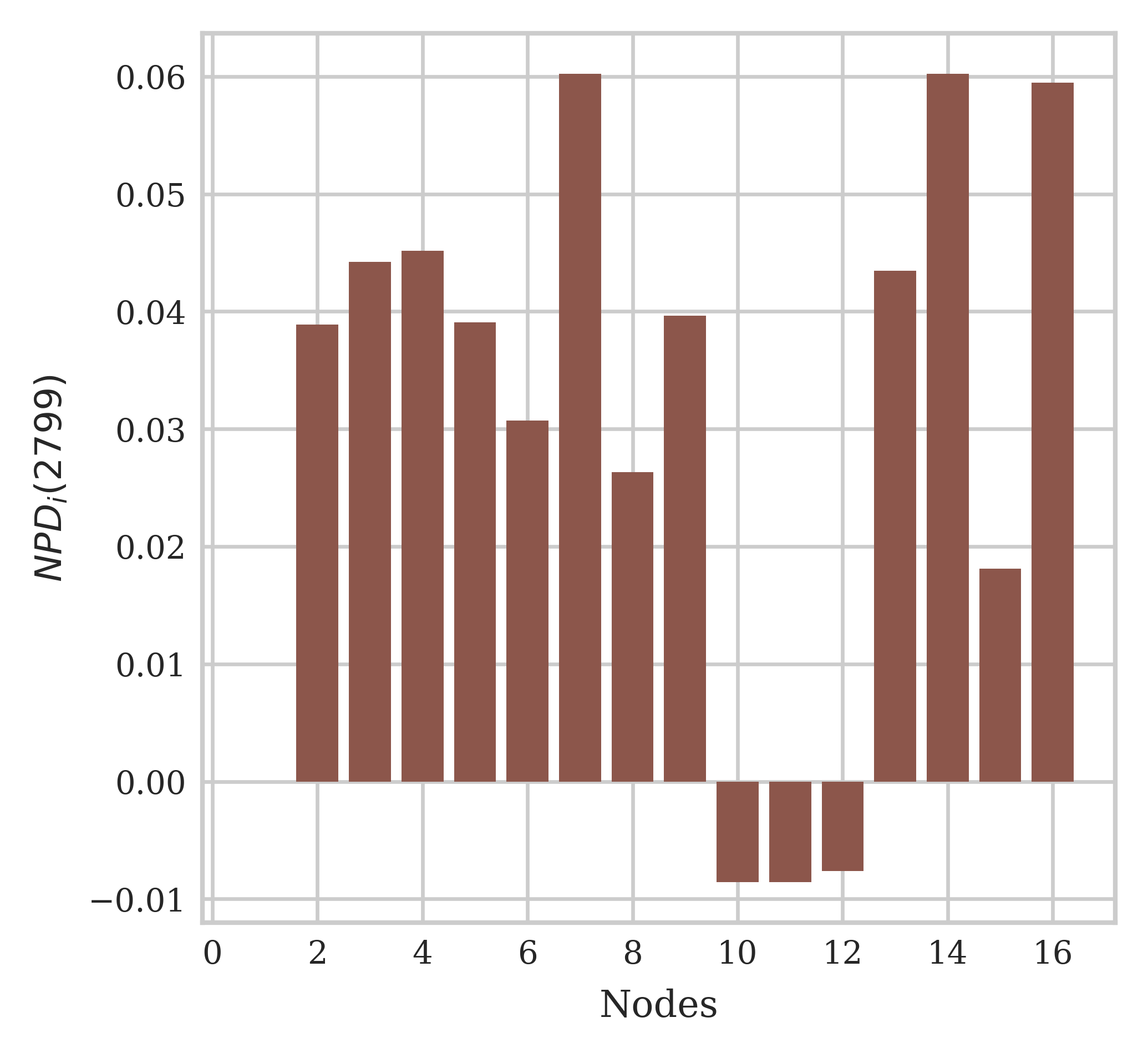}
		\vspace{-0.6cm}
		\caption{}
		\label{fig:Delayspread_ana}
		\end{subfigure}
		\begin{subfigure}[t]{0.45\columnwidth}
		\centering
		\includegraphics[width=\textwidth]{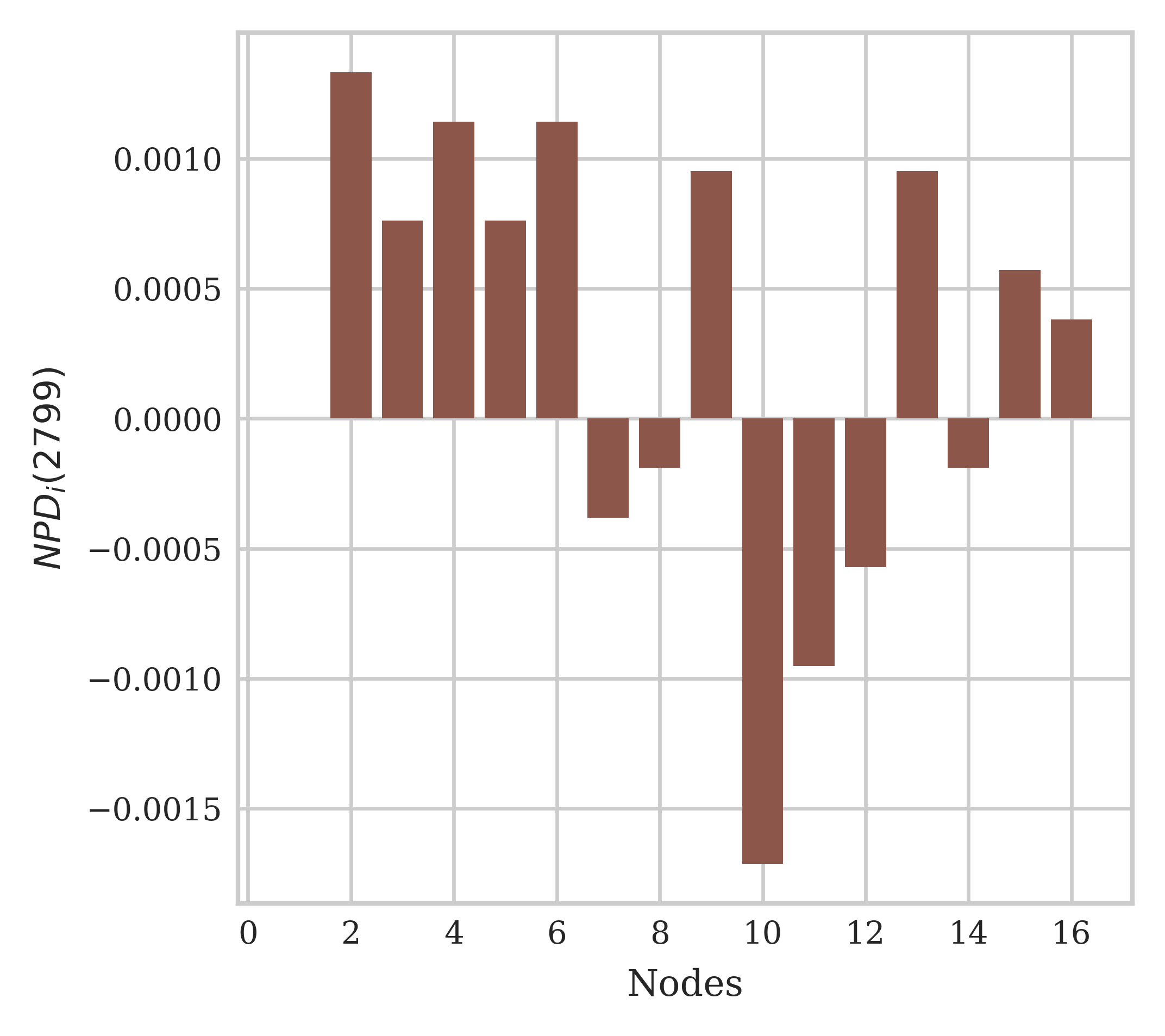}
		\vspace{-0.6cm}
		\caption{}
		\label{fig:Delayspread_dnn}
		\end{subfigure}
			\vspace{-0.3cm}
				\setlength{\belowcaptionskip}{-5pt}
		\caption{Normalized phase difference profile for (a) The classical algorithm and (b)  \ac{dasa}.}
		\label{fig:Delayspread_aftersynch}
		\vspace*{-0.0cm}
	\end{figure}

We further compare the performance of both schemes by observing  the \ac{npd}, defined as the difference between the clock phases at the nodes and the clock phase at node $1$, normalized to the mean period, denoted $T_{c}(k)$. Thus, the \ac{npd} for node $i$ at time $k$ is defined as:
	\begin{eqnarray}
	    \mbox{NPD}_i(k) & = & {\big(\stampi(k)-\phi_1(k)\big)}/{T_{c}(k)},
	    \label{eqn:npd}\\
	    \mbox{NPD range}(k) & \triangleq &  \mathop{\max}\limits _{i\in \mI_N} \ac{npd}_i(k) - \mathop{\min}\limits _{i\in \mI_N} \ac{npd}_i(k).
	    \label{eqn:NPDrange}
	\end{eqnarray}
	where $T_c(k)$ depends on the tested algorithm: For the classic algorithm,  the \ac{npd} is computed w.r.t.  its converged period, denoted $T_c(k)\equiv T_{c,ANA}(k)$, and for the \ac{dasa}   the \ac{npd} is computed w.r.t $T_c(k)\equiv T_{c,DNN}(k)$. 
	The \ac{npd} values for both schemes at $k=2799$ is depicted in Fig. \ref{fig:Delayspread_aftersynch}, and the mean and \ac{std} of  $\mbox{NPD}_i(k)$ over all $i\in\mI_N$ at $k=2799$ are  summarized in  Table \ref{tab:performance_table}.  
	From Fig. \ref{fig:Delayspread_ana} it is observed that the \ac{npd} of analytic algorithm spans a range of $7\%$ of the clock period, with a mean \ac{npd} value of $3\%$,   while the \ac{dasa},  depicted in Fig. \ref{fig:Delayspread_dnn}, achieves  an \ac{npd} range of  $0.35\%$ and a mean \ac{npd} of $0.025\%$. It thus follows that the \ac{dasa} achieves an  improvement by factor of 28 in the standard deviation of the \ac{npd}  and by a factor of $150$ \black in the mean \ac{npd}. 
	We observe from the table that both schemes achieve frequency synchronization, yet only the \ac{dnn}-aided network achieves full and accurate synchronization. In the subsequent simulations we test the robustness of the \ac{dnn}-based scheme to initial clock phase and clock frequency values, and to node mobility, as well as characterize the  performance attained when training is done offline.	

\begin{table}
\vspace{0.0cm}
\centering
\caption{Performance summary for the classic  algorithm and the proposed \ac{dasa}, at  $k=2799$.\label{tab:performance_table}}
\vspace{-0.2cm}
\resizebox{0.55\columnwidth}{!}{
\begin{tabular}{ |l|c|c| } 
 \hline
       & Classical Algorithm & \ac{dasa} \\
 \hline
 Mean period &  $0.00500208$ &  $0.00500774$ \black\\ 
 STD of the period &  $5.4715 e^{-7}$ &  $<10^{-10}$ \\ 
 Mean \ac{npd}    &  $3.0052 e^{-2}$ &    $2.4995 e^{-4}$\\
 STD of \ac{npd}  &  $2.3738 e^{-2}$ &    $8.3749 e^{-4}$ \\
\hline 
\end{tabular}%
}
\vspace{-0.8cm}
\end{table}
\color{black}

\vspace{-0.6cm}

\subsection{Robustness to Clock Phase and Frequency Resets}
\label{subsec:Online_training} 
\vspace{-0.2cm}

In the section we test the robustness of \ac{dasa} to  clock frequency and phase resets during the free-run operation. In the experiments, we first let the nodes learn their \ac{dnn} networks' parameters, $\myVec{\theta}_i$, $i\in\mI_N$, in an unsupervised manner, as described in Section \ref{subsec:training}. 
Then, \acp{dnn}' parameters at the nodes remain fixed, while clock resets are applied. Performance in terms of both speed of convergence after a clock reset and the ability to restore full network clock synchronization after a reset are presented for both  \ac{dasa} and the classic algorithm. 

\begin{figure}
		\centering
		\begin{subfigure}[t]{0.43\columnwidth}
		\centering
		\includegraphics[width=\textwidth]{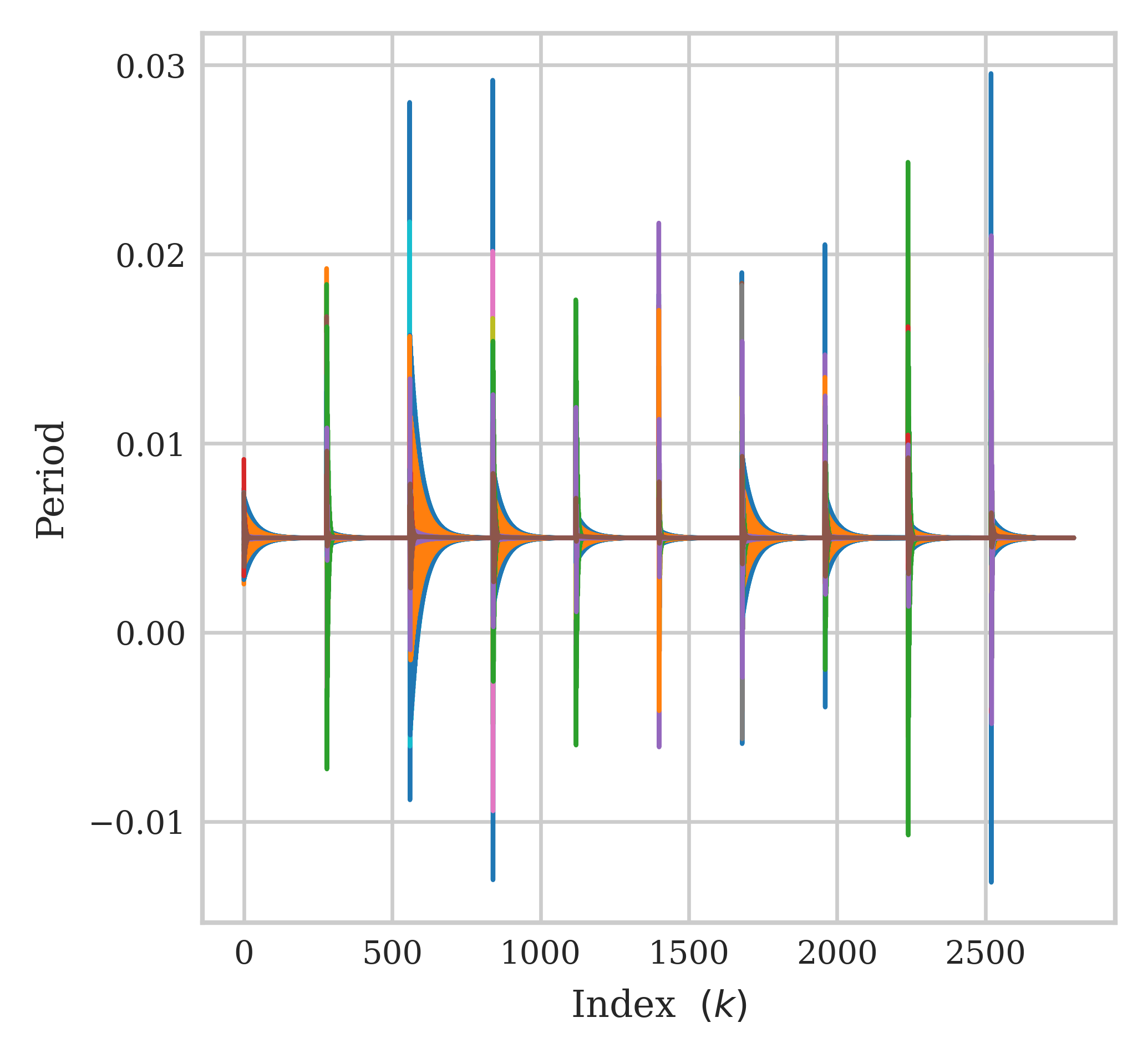}
		\vspace{-0.6cm}
		\caption{}
		\end{subfigure}
		\quad
		\begin{subfigure}[t]{0.43\columnwidth}
		\centering
		\includegraphics[width=\textwidth]{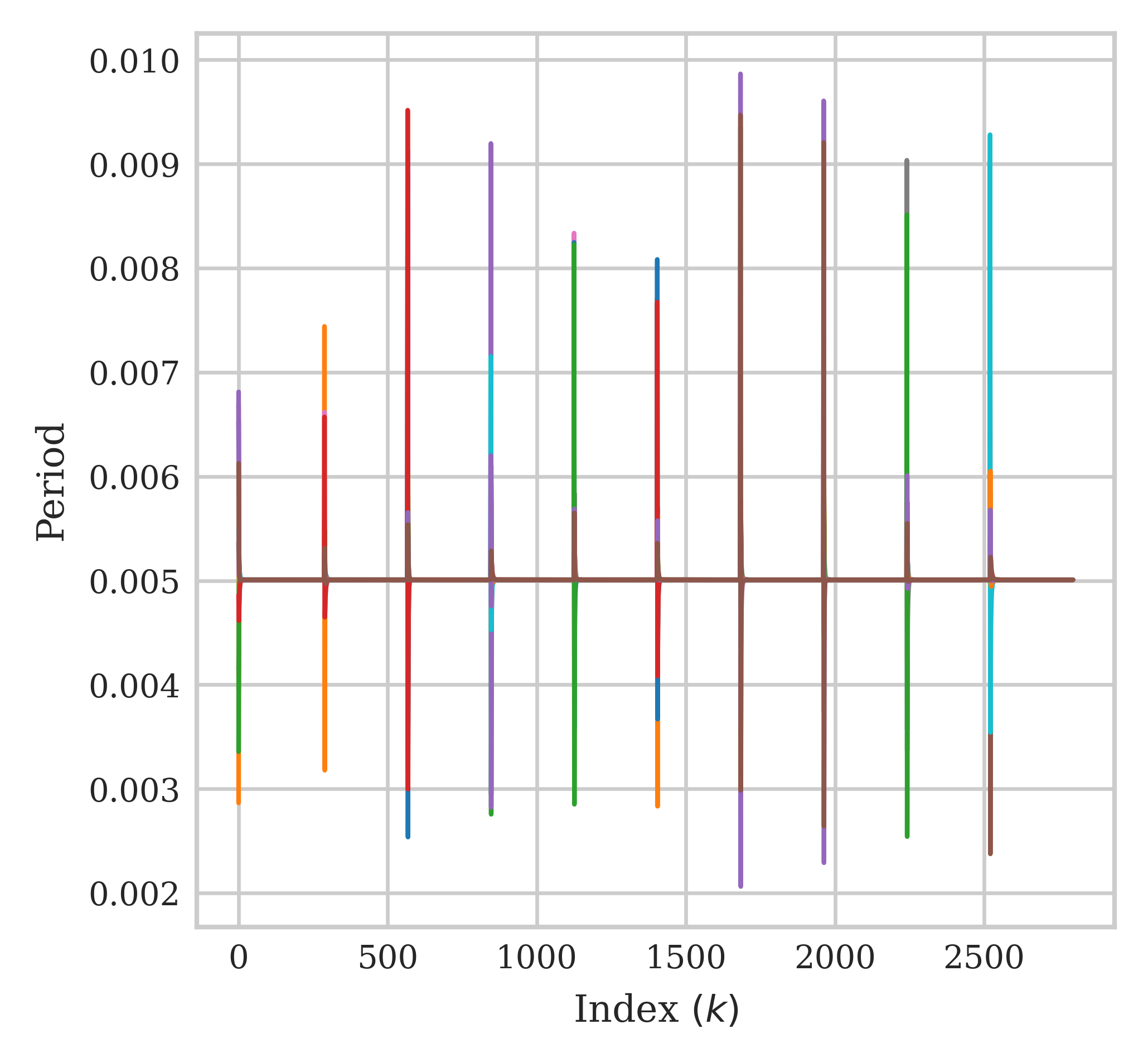}
				\vspace{-0.6cm}
		\caption{}
		\end{subfigure}
				\vspace{-0.3cm}
		\caption{Convergence of clock frequencies after subsequent clock phase and frequency resets over time index for (a) The classic algorithm; and (b)  \ac{dasa}.}
		\label{fig:periodconv_clkfreqreset}
	\end{figure}

\begin{figure}
		\centering
		\begin{subfigure}[t]{0.43\columnwidth}
		\centering
		\includegraphics[width=\textwidth]{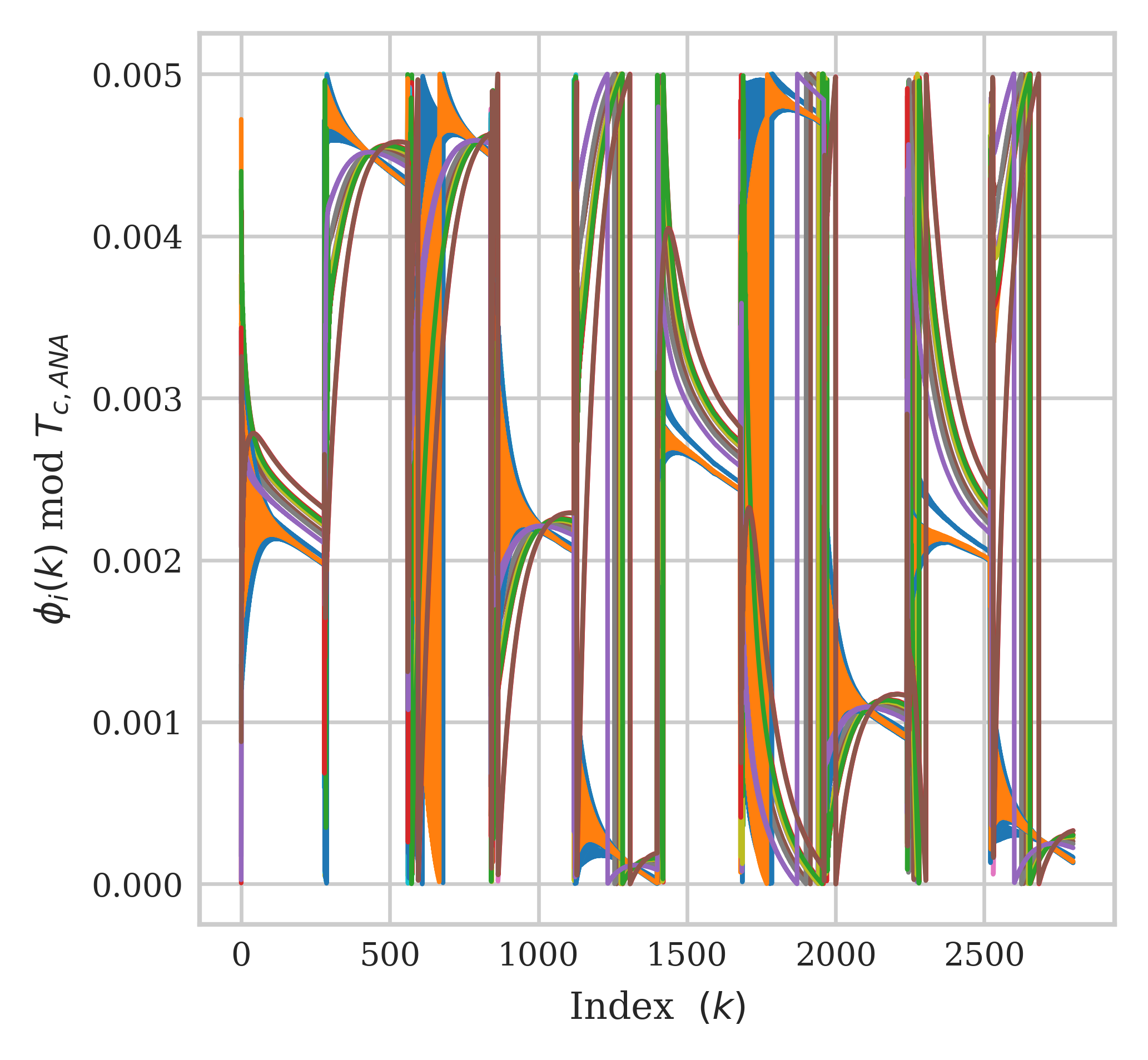}
				\vspace{-0.6cm}
		\caption{}
		
		\end{subfigure}
		\quad
		\begin{subfigure}[t]{0.43\columnwidth}
		\centering
		\includegraphics[width=\textwidth]{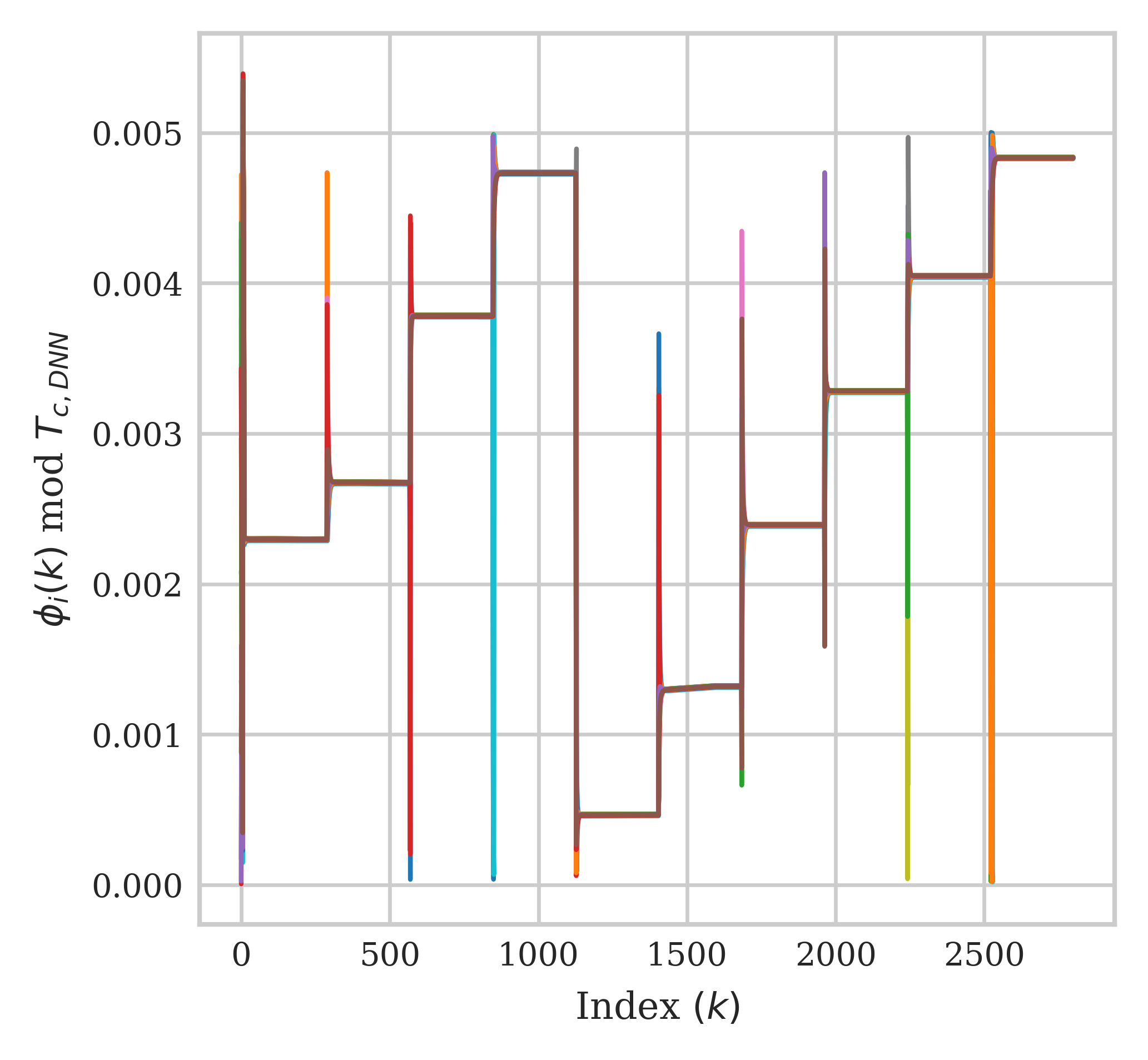}
				\vspace{-0.6cm}
		\caption{}
		\end{subfigure}
				\vspace{-0.3cm}
		\caption{Modulus of clock phases after subsequent clock phase and period resets over time index for (a) The classic algorithm; and (b)  \ac{dasa}.}
		\label{fig:Modulusconv_clkfreqreset}
					\vspace{-0.2cm}
	\end{figure}

In the  experiment, {\em both} the frequencies and the phases of $30\%$ of the nodes were randomly reset, according to the random distributions detailed in Section \ref{sec:Preliminaries}, periodically  every $280$ time instants.
The resulting clock periods and clock phases for all the nodes in the network are depicted in Figs. \ref{fig:periodconv_clkfreqreset} 
and \ref{fig:Modulusconv_clkfreqreset}, 
respectively, for the classic algorithm as well as for  \ac{dasa}.  It is observed from Fig. \ref{fig:periodconv_clkfreqreset} 
that both the classic algorithm and  \ac{dasa} are able to restore frequency synchronization, yet the  proposed \ac{dasa} is able to {\em instantly} restore frequency synchronization. We observe from Fig. \ref{fig:Modulusconv_clkfreqreset} 
that the slow frequency synchronization of the classic algorithm induces  slow phase synchronization, which is not completed before the next reset occurs, while the newly proposed \ac{dasa} instantly restores phase synchronization.  Is is observed in   Fig. \ref{fig:Modulusconv_clkfreqreset} that the converged (i.e. steady state) phases of  \ac{dasa} after clock resets are different, yet we clarify that this  has no impact on communications network's performance as all the nodes converge to the {\em same phase} within the converged period.
It is observed that our proposed \ac{dasa} is able to {\em instantly restore both the clock frequency and clock phase synchronization} (namely, full synchronization), while the classic algorithm  requires longer convergence times, and  its phases do not complete the convergence process before the next clock reset is applied.

\begin{figure}[h]
		\centering
        \begin{subfigure}[t]{0.43\columnwidth}
		\centering
		\includegraphics[width=\textwidth]{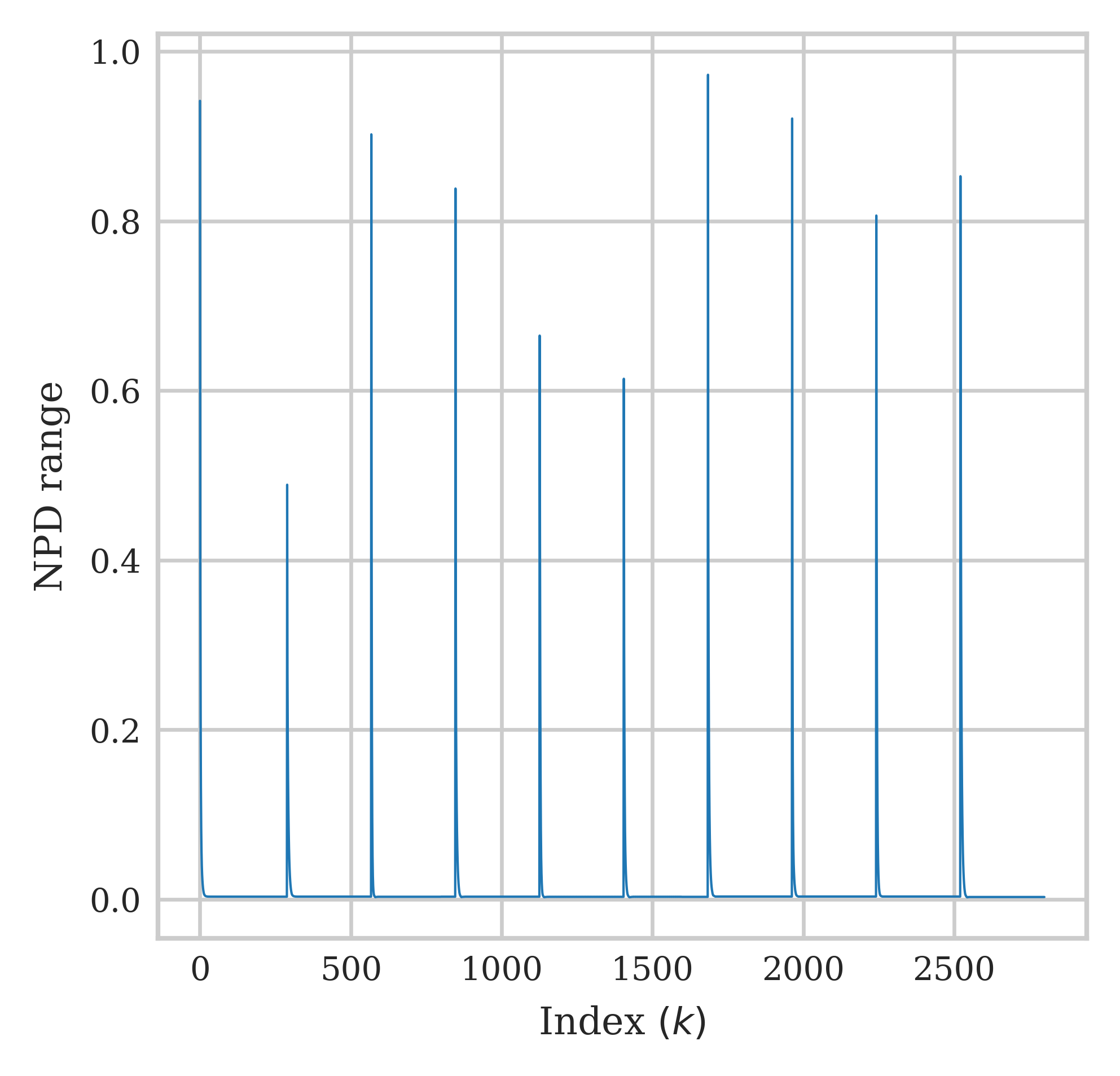}
				\vspace{-0.6cm}
		\caption{}
		\label{fig:delayspread_clockperiod_reset}
		\end{subfigure}
		\begin{subfigure}[t]{0.43\columnwidth}
		\centering
		\includegraphics[width=1.05\textwidth]{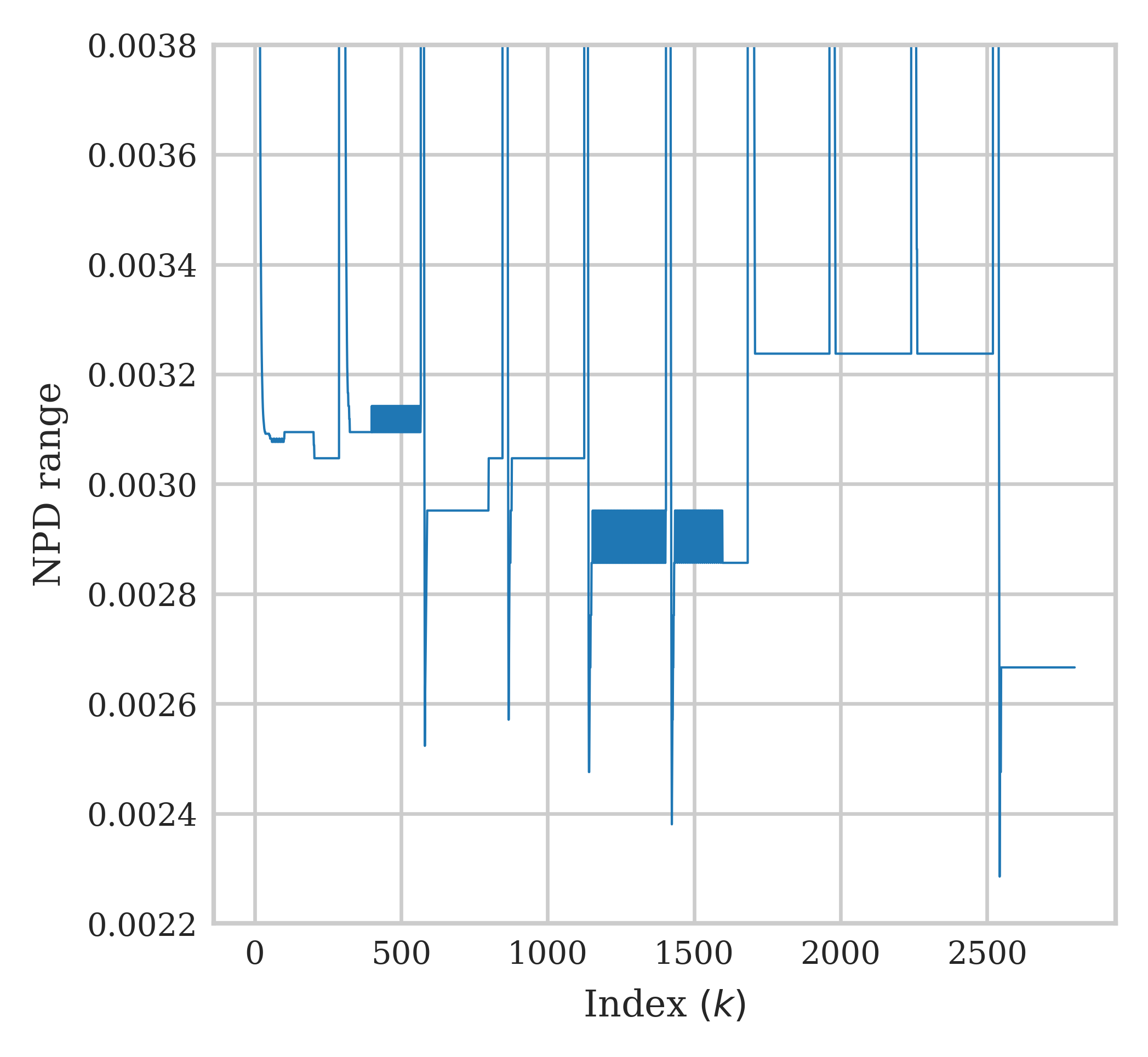}
				\vspace{-0.6cm}
		\caption{}
		\label{fig:zoomeddelayspread_clockperiod_reset}
		\end{subfigure}
		\vspace{-0.3cm}
			\setlength{\belowcaptionskip}{-10pt}
		\caption{ \ac{npd} range  over the time index $k$ with  subsequent clock phase and period resets over time index and (a) Overall (b) Closeup.}
		\label{fig:Delayspread_clck_clckPer}
	\end{figure}

Next, we focused on the \ac{npd} maintained by  \ac{dasa} during the clock resets. To that aim we plot  in Fig. \ref{fig:Delayspread_clck_clckPer} the \ac{npd} range, i.e., the difference between the maximal \ac{npd} and the minimal \ac{npd}, achieved by \ac{dasa} when both clock phase and period resets are applied. 
The overall \ac{npd} is depicted in Fig. \ref{fig:delayspread_clockperiod_reset}, where a zoom on the smaller value range, corresponding to the converged state is depicted in Fig. 
\ref{fig:zoomeddelayspread_clockperiod_reset}. Comparing Fig. \ref{fig:zoomeddelayspread_clockperiod_reset} and Fig. \ref{fig:Delayspread_dnn} it is observed that after clock resets  \ac{dasa} achieves the {\em same \ac{npd}} as in the initial operation after training, namely, {\em clock resets do not degrade the small \ac{npd} achieved by  \ac{dasa}}.
The experiments demonstrate that the proposed \ac{dasa} is  able to facilitate nearly uninterrupted clock phase synchronization, also in presence of random clock resets.  
 These experiments clearly show that {\em  \ac{dasa} is robust to the initial phase and has an outstanding ability to recover from clock phase and frequency variations}, which may occur due to e.g., clock temperature changes.

\vspace{-0.05cm}
	
\subsection{Testing DASA Synchronization Performance with Mobile Nodes}
\label{Online_mobility}

\begin{figure}
		\centering
		\begin{subfigure}[t]{0.43\columnwidth}
		\centering
		\includegraphics[width=\textwidth]{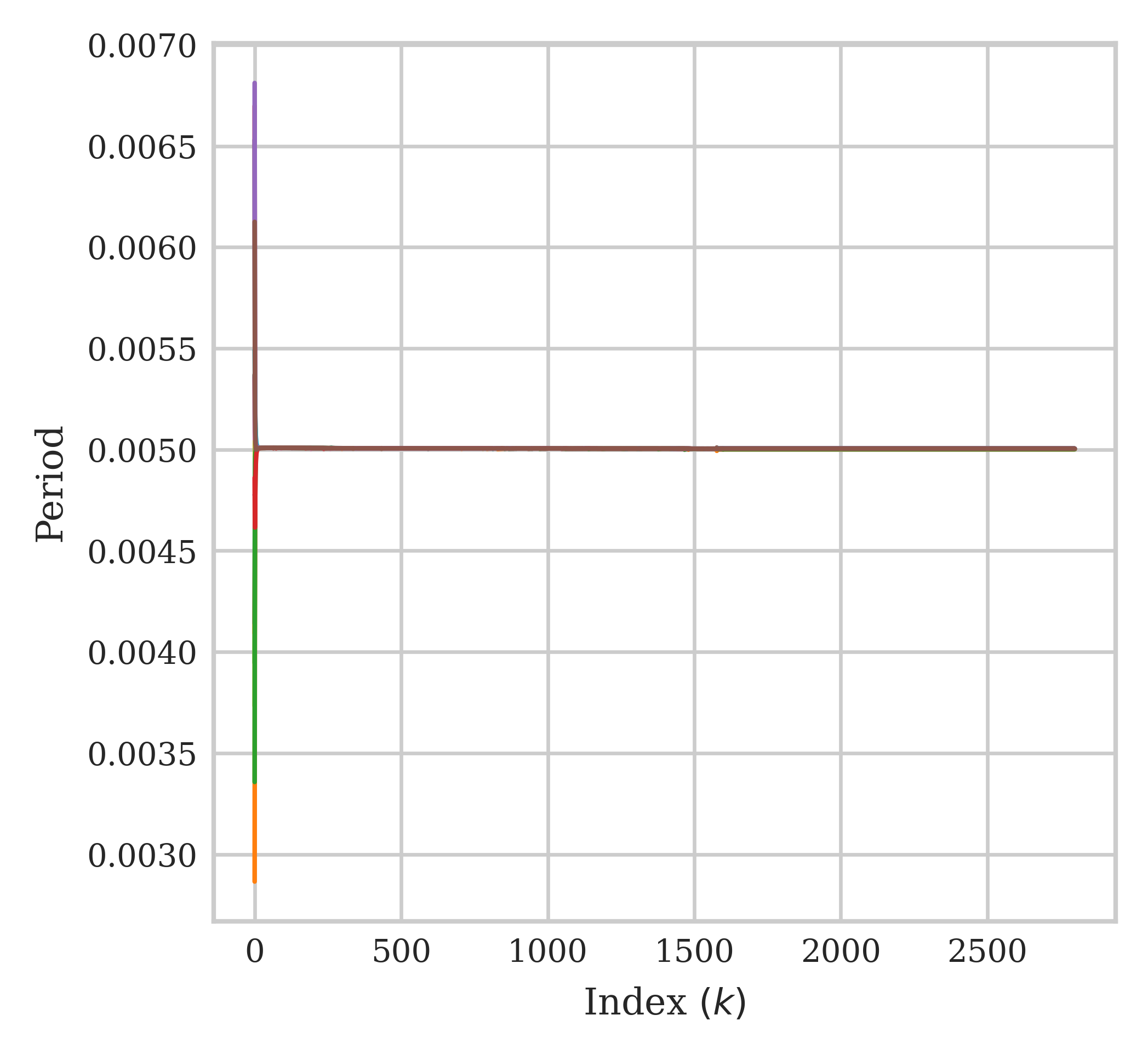}
		\vspace{-0.8cm}
		\caption{}
		\label{fig:frequency_walk_alltimes}
		\end{subfigure}		
		\begin{subfigure}[t]{0.43\columnwidth}
		\centering
		\includegraphics[width=\textwidth]{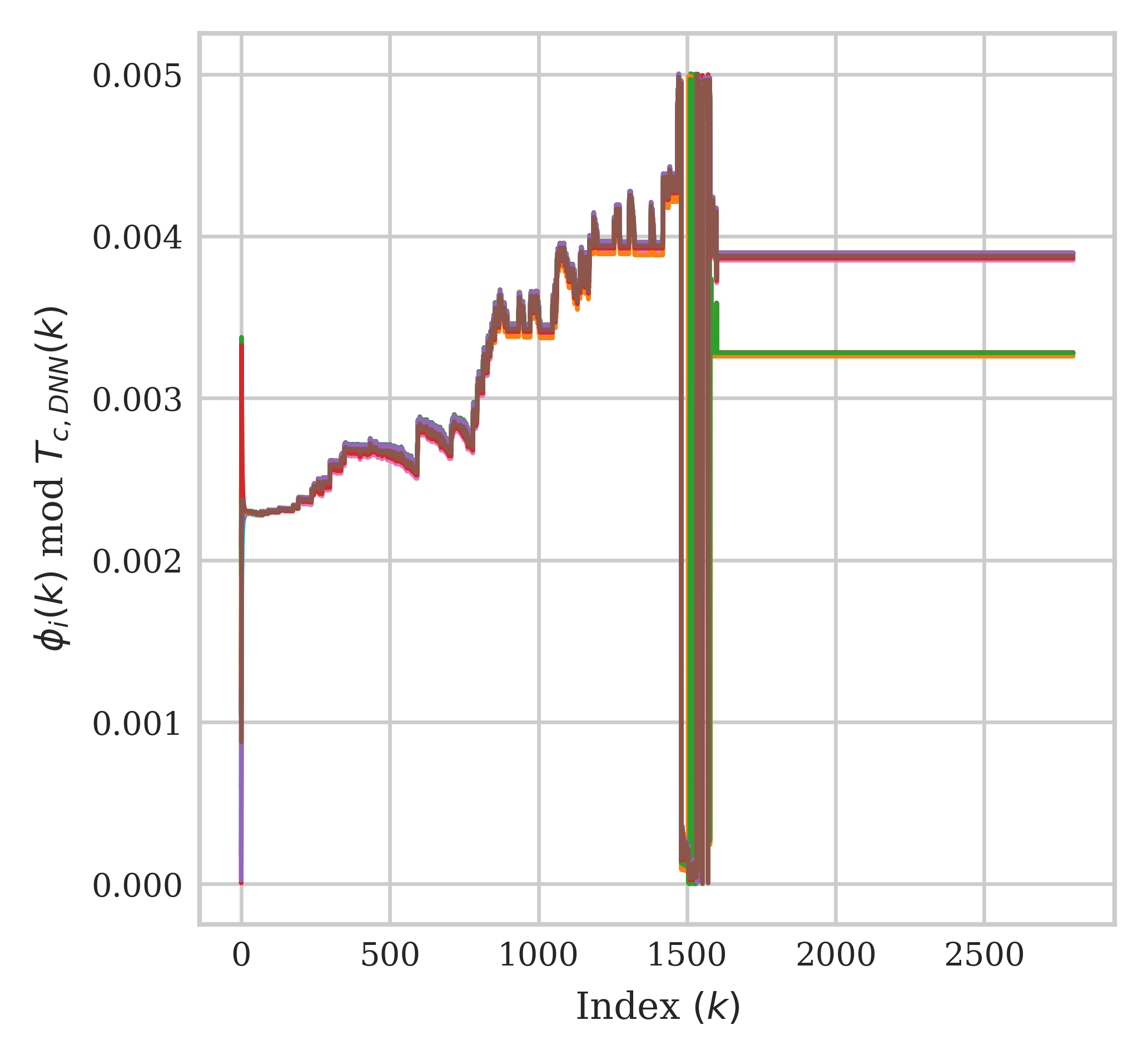}
		\vspace{-0.8cm}
		\caption{}
		\label{fig:modulophase_walk_alltimes}
		\end{subfigure}

        \vspace{-0.3cm}
		\caption{Clocks' periods and phases with mobility of $30$\% of the nodes at inference 
		w.r.t. the time indices: (a) Clock periods and (b) Clock phases modulo the instantaneous period, $T_{c,DNN}(k)$. }
		\label{fig:FreqmoduloPhase_walk_alltimes}
		\vspace{-0.0cm}
	\end{figure}

In this subsection we test synchronization performance when some of the nodes are mobile. We let the \acp{dnn} at the nodes converge for a stationary scenario (i.e., online training) and then examine synchronization performance when a random subset of  $30\%$ of the nodes, selected uniformly, 
begins moving at a fixed speed, with each mobile given an angular direction.
Note that as the nodes move, the received signal powers from and at the moving nodes vary and the received signals at some of the nodes for certain links may fall below the receive threshold, which for the current setup is set to $-114$ dBm. This situation is implemented for such links by setting both the phase difference and the received power to zero. Naturally, this assignment should have a negative impact on synchronization accuracy. 
In the first experiment, in order to demonstrate the situation of node clustering, the moving nodes were all given the same direction of $95^{\circ}$ and the moving speed was se such that at the end of the simulation each node has traversed $20$ [Km]. Fig. \ref{fig:FreqmoduloPhase_walk_alltimes} depicts  the clock periods and clock phases modulo the instantaneous mean period, $T_{c,DNN}(k)$.
It is observed from Fig. \ref{fig:frequency_walk_alltimes} that frequency synchronization is largely  maintained also when nodes are mobile, yet, from Fig. \ref{fig:modulophase_walk_alltimes} we observe a  slow drift in the phase modulo  $T_{c,DNN}(k)$, which implies that the period slightly varies as the nodes move. It is also noted that despite the  phase drift, the nodes are able to maintain close phase values up to a certain time (in this simulation it is time index $1576$, corresponding to a displacement of $10.9$ [Km]), after which the phases split into two separate branches, one consisting of the five mobile i.e., nodes $1$, $2$, $3$, $11$, and $12$, and the second corresponding to the stationary nodes.
Checking the connectivity graph for this scenario, it was discovered that at this time index, the network splits into two disconnected sub-networks. Observe that at each sub-network the nodes maintain phase synchronization among themselves. 

\begin{figure}[t]
		\centering
		\includegraphics[width=0.55\textwidth]{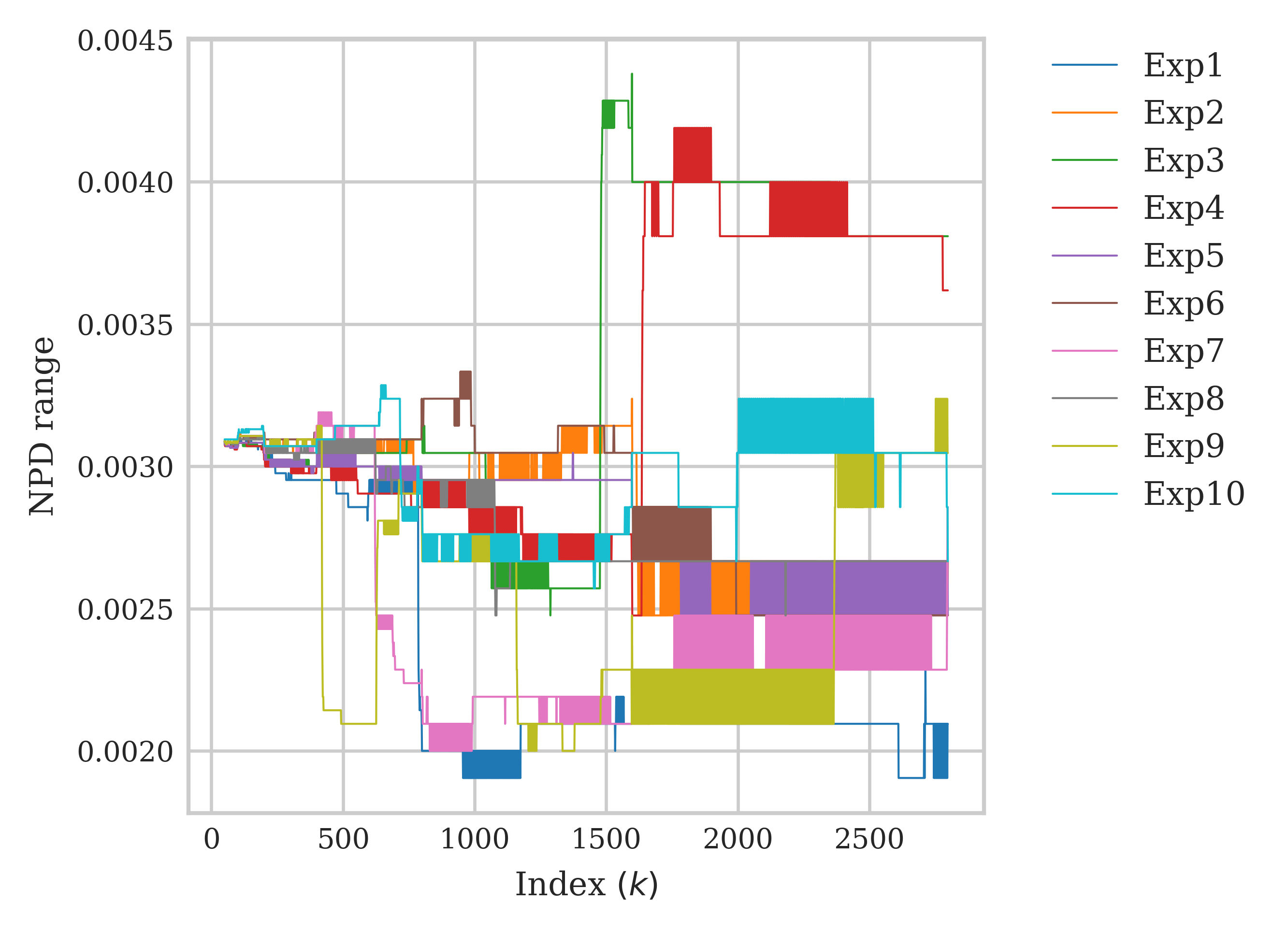}
    \vspace{-0.3cm}
		\caption{
		\ac{npd} range for $10$ experiments, with online training, where each mobile node moves at a different direction and traverses $1$ [Km] during the simulation. }
		\label{fig:NPD_walk_allexperimentsDiffDirections}
	
	\end{figure}

Lastly, we  take a closer look at the \ac{npd} performance before network splitting occurs. To that aim, we carried out a new set of $10$ mobility experiments, where the nodes are moving as a speed of $250$ [Km/h], 
such that at the last time index, $k=2799$, the distance each mobile node traversed was $1$ [Km].  The direction of each moving node is randomly and uniformly selected over $[0,2\pi]$. Fig. \ref{fig:NPD_walk_allexperimentsDiffDirections} depicts the evolution of the \ac{npd} range w.r.t $T_{c,DNN}(k)$, for all $10$ experiments. It is observed that the \ac{npd} range maintains a very small value, and in fact, in all experiments the \ac{npd} range is increased by a factor smaller than $2$ after a $1$ [Km] displacement.
Hence, we obtain that  \ac{dasa}  exhibits a graceful degradation when the node locations vary, and very good accuracy is maintained also for a large displacement of $1$ [Km]. In fact, it seems that  as long as movement does not result in disconnected clusters,  \ac{dasa}  generally maintains network synchronization. This demonstrates the robustness of  \ac{dasa} to the topology of the network. 
This point is of major importance as it implies that moderate node displacements do not require retraining of the network. As a last note in this context, then, based on the mobility experiments and the clock reset experiments we conclude that  the optimal weights are  dependent on the relative locations of the nodes in the network and only weakly dependent on the clock frequencies and phases.

\vspace{-0.2cm}
\subsection{Offline Training}
\label{subsec:Online_trainingMobile}

Next, we examine the possibility of training the \acp{dnn} offline over multiple topologies, instead of training online at startup for the actual topology.
To implement offline training, we generate a set of $N_{top}$  network topologies, each generated over the same network area of $10$~[km] $\times$ $10$ [km]. For each topology instance, the individual nodes acquire simulated training data consisting of  $N_T$ time intervals, where at each interval, each node collects $N-1$ inputs. Each input consists of a pair of receive time and received power levels as specified in Eqn. \eqref{eqn:dataset}. The training data for the $i$'th node over the entire samples of network topologies is denoted as
 $   \mySet{D}_i^{N_{top}}=\left\{\mySet{D}_{i,n_{top}}\right\}_{n_{top}=1}^{N_{top}}$,
where $\mySet{D}_{i,n_{top}}$ denotes the training data at node $i$ for the $n_{top}$'th topology.

In the previous tests we computed the training loss for a single topology, such that the loss is computed over a single batch and only one step of \ac{gd} is applied at each epoch. In contrast, for the offline training  we apply the \ac{mbgd}. In this approach, the set $\mySet{D}_i^{N_{top}}$ is divided into mini batches of a fixed size denoted by $N_{samp}$. At each epoch, the \ac{mbgd} at node $i$ operates as follows:
\begin{enumerate}
    \item \label{step1} A mini-batch is selected for training the node's \ac{dnn} in a sequential order.
    \item Estimate the average loss over the mini batch topologies, where the loss for the $n_{top}$'th topology, denoted by $\mySet{L}_{\mySet{D}_{i,n_{top}}}(\myVec{\theta}_i)$  is obtained via Eqn. \eqref{eqn:loss1}.
    \item  \label{step3} Compute the gradient and update the \ac{dnn}'s weights using computed gradient.
    \item Repeat steps \ref{step1}-\ref{step3} for all mini-batches.
\end{enumerate}
    The  mini batch training procedure is summarized in 
    Algorithm \ref{alg:UnsupLocTrainingOffline_alg}. 
    In the numerical evaluation we used $N_{top}=1000$ topologies, the mini-batch size was set to $N_{samp}=10$; hence, there are $N_{batch}=N_{top}/N_{samples}=100$ mini-batches. For the considered numerical evaluation with \ac{mbgd},  setting  $E=3$ epochs was found sufficient to achieve convergence. 
    
\setlength{\textfloatsep}{10pt}

\RestyleAlgo{ruled}
\begin{algorithm}
    \caption{Unsupervised Offline Local Training at Node $i$
    }\label{alg:UnsupLocTrainingOffline_alg} 
    \KwData{Data set $\mySet{D}_i^{N_{top}}$, learning rate $\mu$, initial weights $\myVec{\theta}_i$,  period $T_i$, number of epochs $E$}
    \For{${\rm epoch}=1$ to $E$}{
    Randomly shuffle $\mySet{D}_i^{N_{top}}$\;
    { \For{${\rm batch}=0$ to $N_{batch}-1$}{
    \For{$k=1$ to $N_T$}{
    \textbf{Forward pass} $\left\{\left\{\big(t_{i,j}(k), P_{i,j}(k)\big) \right\}_{j=1, j\neq i}^{N} \right\}_{n_{top}={\rm batch}\cdot N_{samp}+1}^{({\rm batch}+1)\cdot N_{samp}+1}\in \mySet{D}_i^{N_{top}}$ to obtain $\left\{\psi_{\myVec{\theta}_{i,n_{top}}}\right\}_{n_{top}={\rm batch}\cdot N_{samp}+1}^{({\rm batch}+1)\cdot N_{samp}+1}$\;  
    \textbf{Evaluate} $\left\{\phi_{i, n_{top}}(k+1)\right\}_{n_{top}={\rm batch}\cdot N_{samp}+1}^{({\rm batch}+1)\cdot N_{samp}+1}$  applying Eqn. \eqref{eqn:clock_analyticalDNN2} for each $n_{top}$;
    }
    \textbf{Compute loss} $\mySet{L}_{\mySet{D}_i}({\rm batch},\myVec{\theta}_i)\triangleq
    \frac{1}{N_{samp}}\mathop{\sum}\limits_{n_{top}={\rm batch}\cdot N_{samp}+1}^{({\rm batch}+1)\cdot N_{samp}+1}\mySet{L}_{\mySet{D}_{i,n_{top}}}(\myVec{\theta}_i)$ via Eqn. \eqref{eqn:loss1}\;
    \textbf{Compute gradient} $\nabla_{\myVec{\theta}_i}\mySet{L}_{\mySet{D}_i}({\rm batch},\myVec{\theta}_i)$ using back propagation through time\;
    \textbf{Update weights} via $\myVec{\theta}_i \leftarrow \myVec{\theta}_i - \mu \cdot \nabla_{\myVec{\theta}_i}\mySet{L}_{\mySet{D}_i}({\rm batch},\myVec{\theta}_i)$.
    }
    }
    }
\end{algorithm}

\begin{figure}[h]
		\centering
		\begin{subfigure}[t]{0.43\columnwidth}
		\centering
		\includegraphics[width=\textwidth]{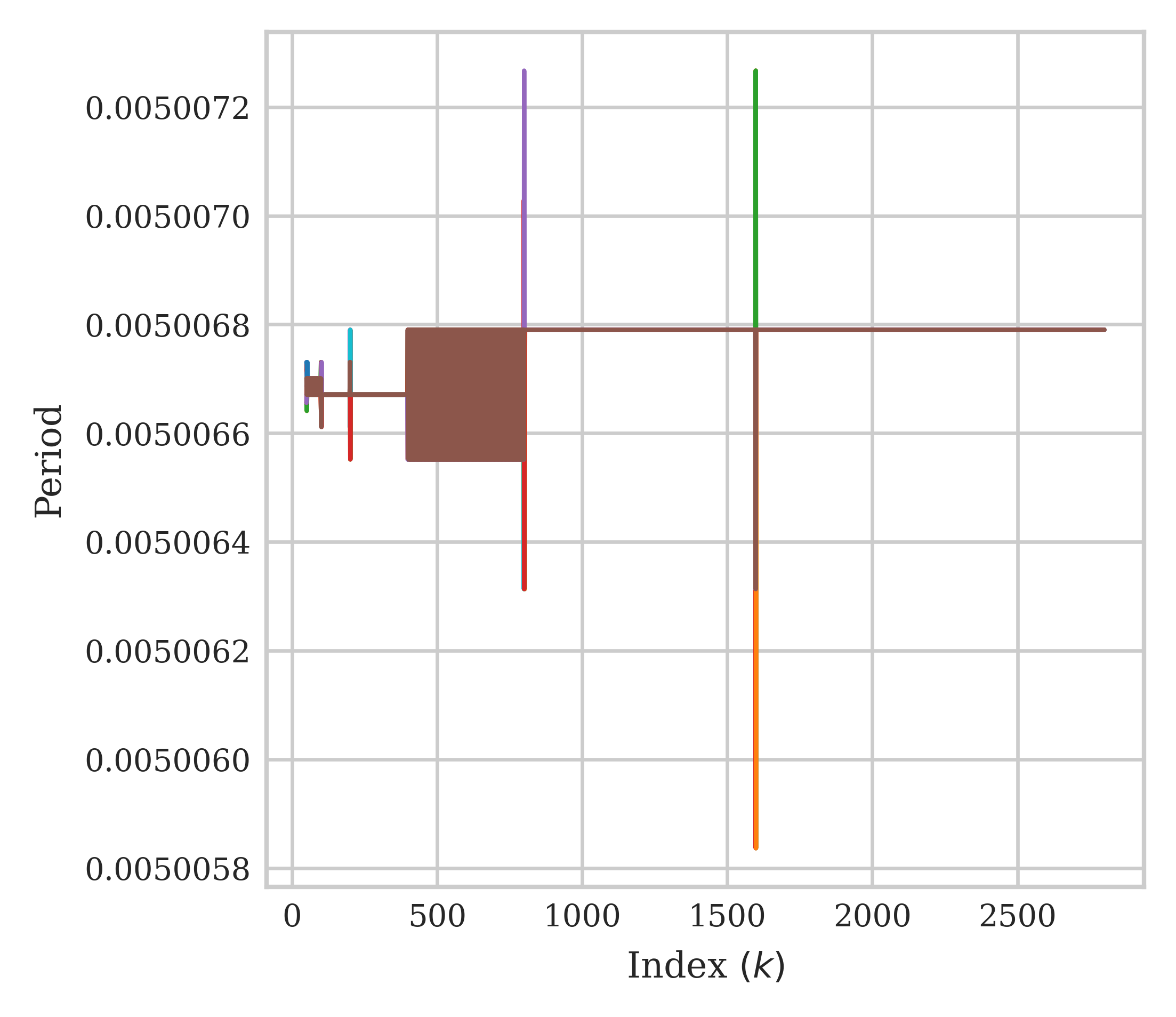}
			\vspace{-0.8cm}
		\caption{}
		\label{fig:frequency_offlineTest}
		\end{subfigure}		
		\begin{subfigure}[t]{0.43\columnwidth}
		\centering
		\includegraphics[width=0.94\textwidth]{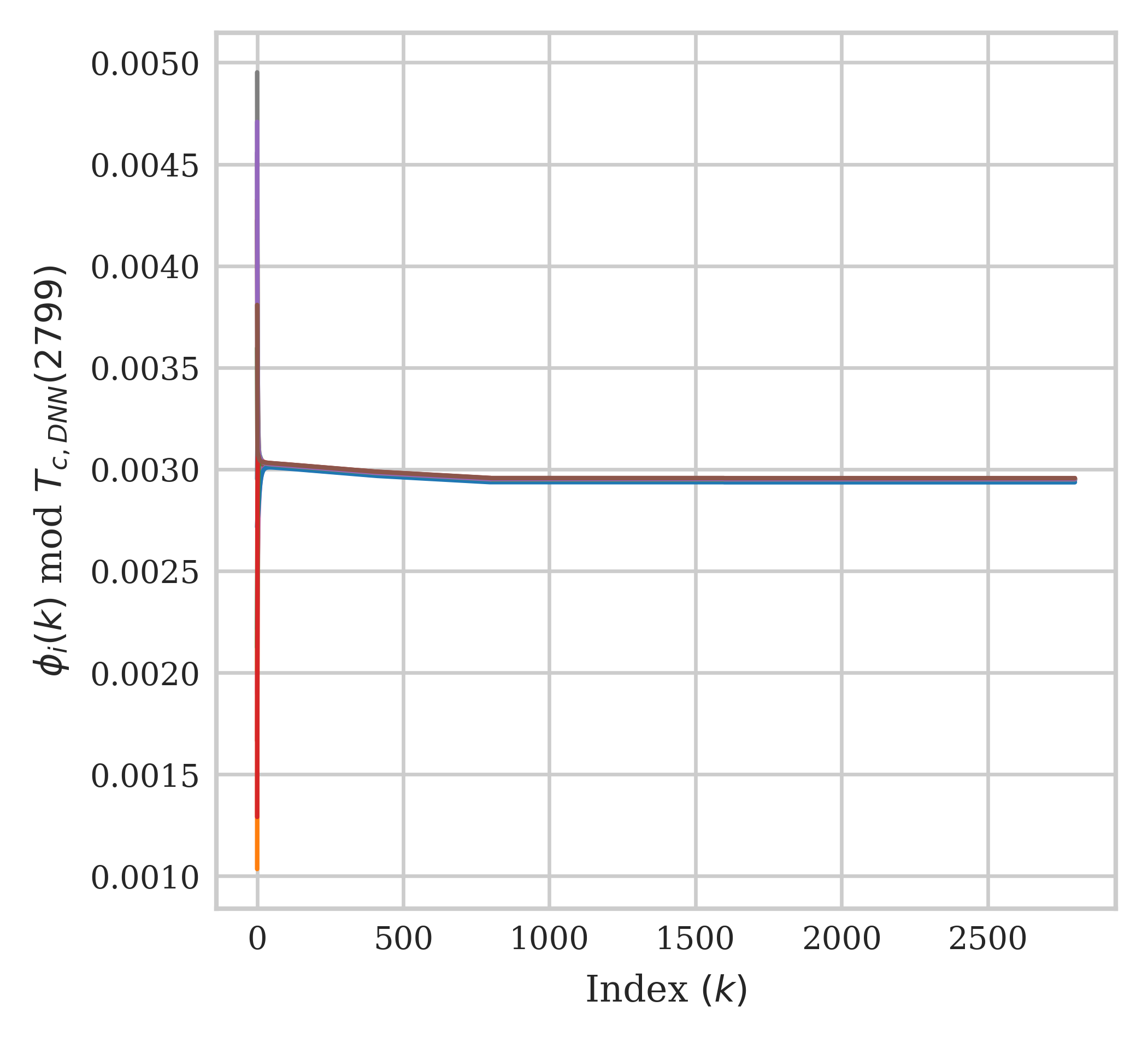}
			\vspace{-0.2cm}
		\caption{}
		\label{fig:modulophase_offlineTest}
		\end{subfigure}
		\begin{subfigure}[t]{0.43\columnwidth}
		\centering
		\includegraphics[width=\textwidth]{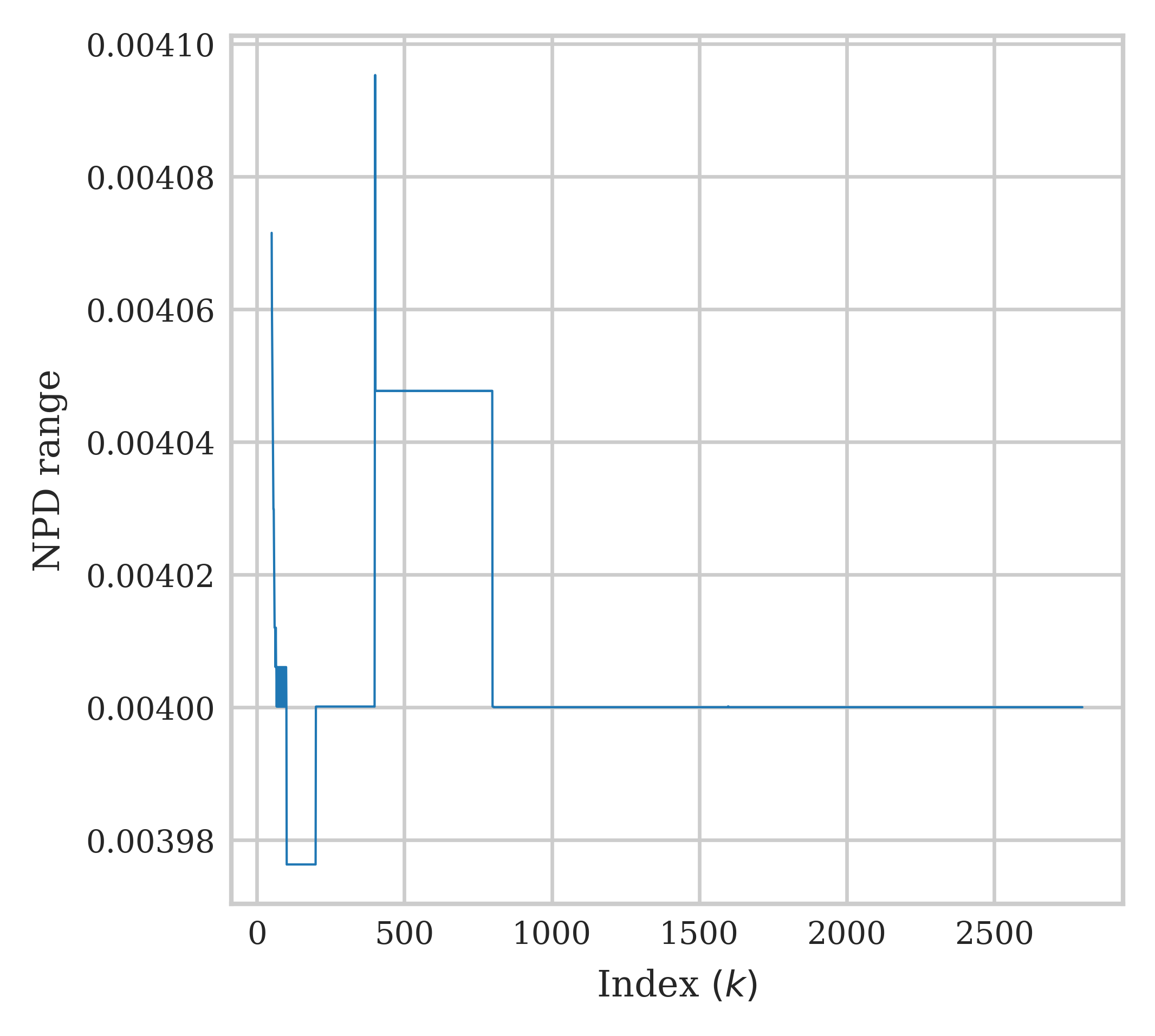}
			\vspace{-0.8cm}
		\caption{}
		\label{fig:NPDrangeOffline}
		\end{subfigure}
		\begin{subfigure}[t]{0.43\columnwidth}
		\centering
		\includegraphics[width=\textwidth]{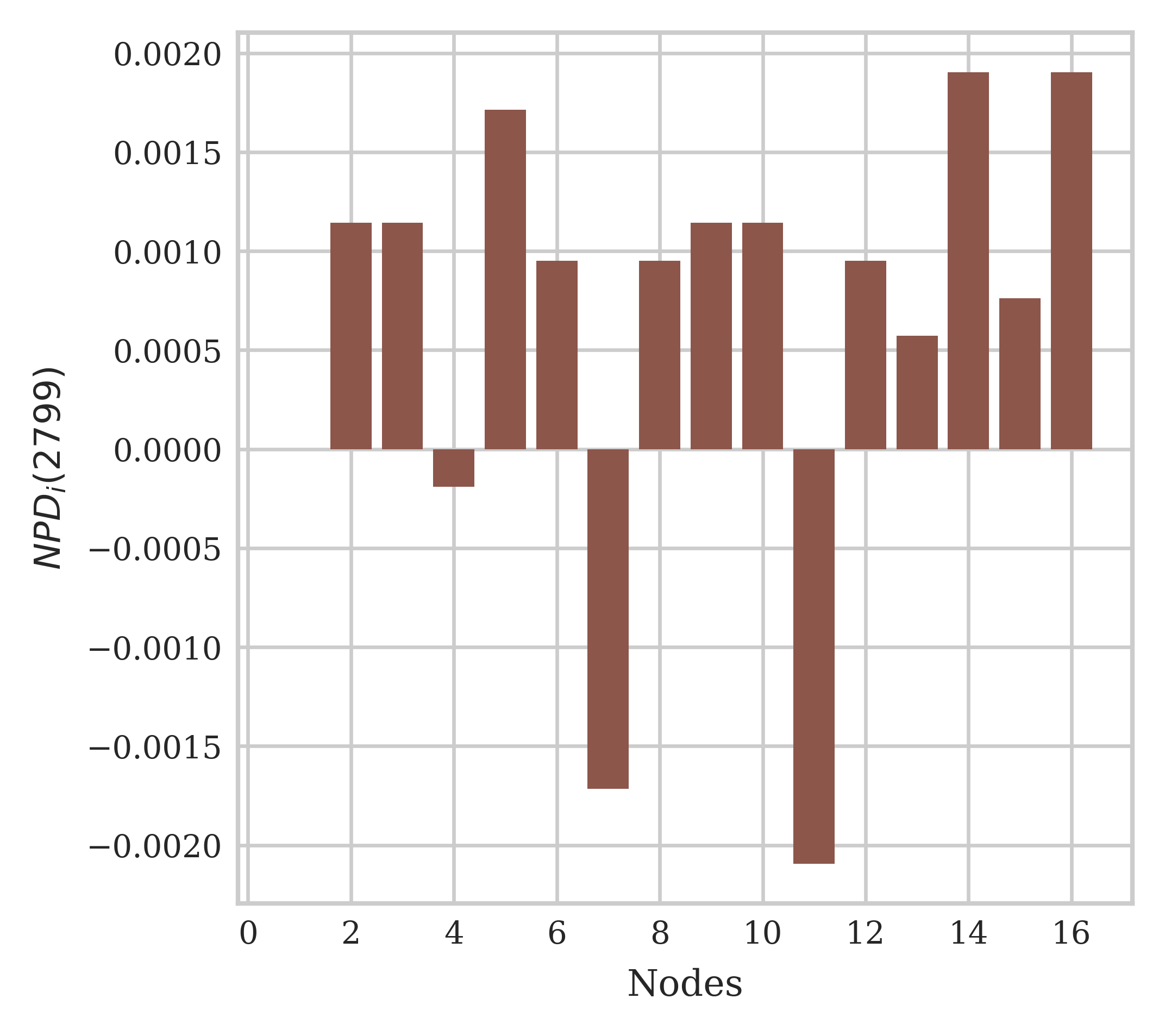}
			\vspace{-0.8cm}
		\caption{}
		\label{fig:NPD_offlineTest}
		\end{subfigure}
		
        \vspace{-0.3cm}
		\caption{Performance of the \ac{dasa} for a test topology after offline training: (a) Clock periods, (b) Clock phases modulo $T_{c,DNN}(2799)$, (c)  NPD range,  and (d) $\big\{NPD_i(2799)\big\}_{i=1}^N6$.}
		\label{fig:FreqmoduloPhase_offlineTest}
\end{figure}

	After the \acp{dnn} have been trained over the set of  $N_{top}=1000$ network topologies,  \ac{dasa} was tested for new topologies not included in the training set. Fig. \ref{fig:FreqmoduloPhase_offlineTest} depicts the results for a test topology sample:  Fig. \ref{fig:frequency_offlineTest} demonstrates the rapid convergence of the clock periods  to a mean synchronized period of $T_{c,DNN}(2799)=0.00500679$. We  observe some fluctuations in the periods of the nodes, however, the amplitudes of these variations are  three orders of magnitude smaller than the mean synchronized period, hence, these variations are rather negligible. Fig. \ref{fig:modulophase_offlineTest} depicts the modulus of the clocks' phases w.r.t. the mean synchronized period $T_{c,DNN}(2799)$. The figure demonstrates that {\em the proposed \ac{dasa} with offline training   achieves full clock synchronization}. Furthermore, its performance is significantly better than the performance of the classical algorithm, as it is robust to propagation delays. 
	Fig. \ref{fig:NPDrangeOffline}, depicts a closeup of the \ac{npd}  range. Observe that  \ac{dasa} achieves an \ac{npd}  range   of $0.41\%$ at the first few time indices and a converged \ac{npd} range of $0.4\%$ at later time indices ($k \ge 850$). 
	Lastly, Fig \ref{fig:NPD_offlineTest} depicts a snapshot of the \ac{npd} values across nodes at time $k=2799$. From the figure, we again note that the \ac{npd} range is $0.4\%$ across the nodes, we also see that the mean value is $0.064\%$. The performance of  \ac{dasa} for this test is summarized in Table \ref{tab:performance_tableOffline}. Comparing with the online training results in Table \ref{tab:performance_table} we note that period accuracy is similar for both scenarios; the main benefit of online training is a smaller \ac{npd}, by a factor of $2.5$, and an \ac{npd} \ac{std} smaller by a factor of $1.3$.
	
\begin{table}[h]
\vspace{0.0cm}
\centering
\caption{Performance summary of the proposed \ac{dasa}, at index k=2799 for a test network topology with offline training setup.\label{tab:performance_tableOffline}}
\resizebox{0.35\columnwidth}{!}{
\begin{tabular}{ |l|c| } 
 \hline
       & \ac{dasa} \\
 \hline
 Mean period &  $0.00500679$ \black\\ 
 STD of the period &  $<10^{-10}$ \\ 
 Mean \ac{npd}   &    $6.4285 e^{-4}$\\
 STD of \ac{npd} &    $1.1124 e^{-3}$ \\
\hline 
\end{tabular}%
}

\end{table}

\begin{figure}[t]
		\centering
		\includegraphics[width=0.55\textwidth]{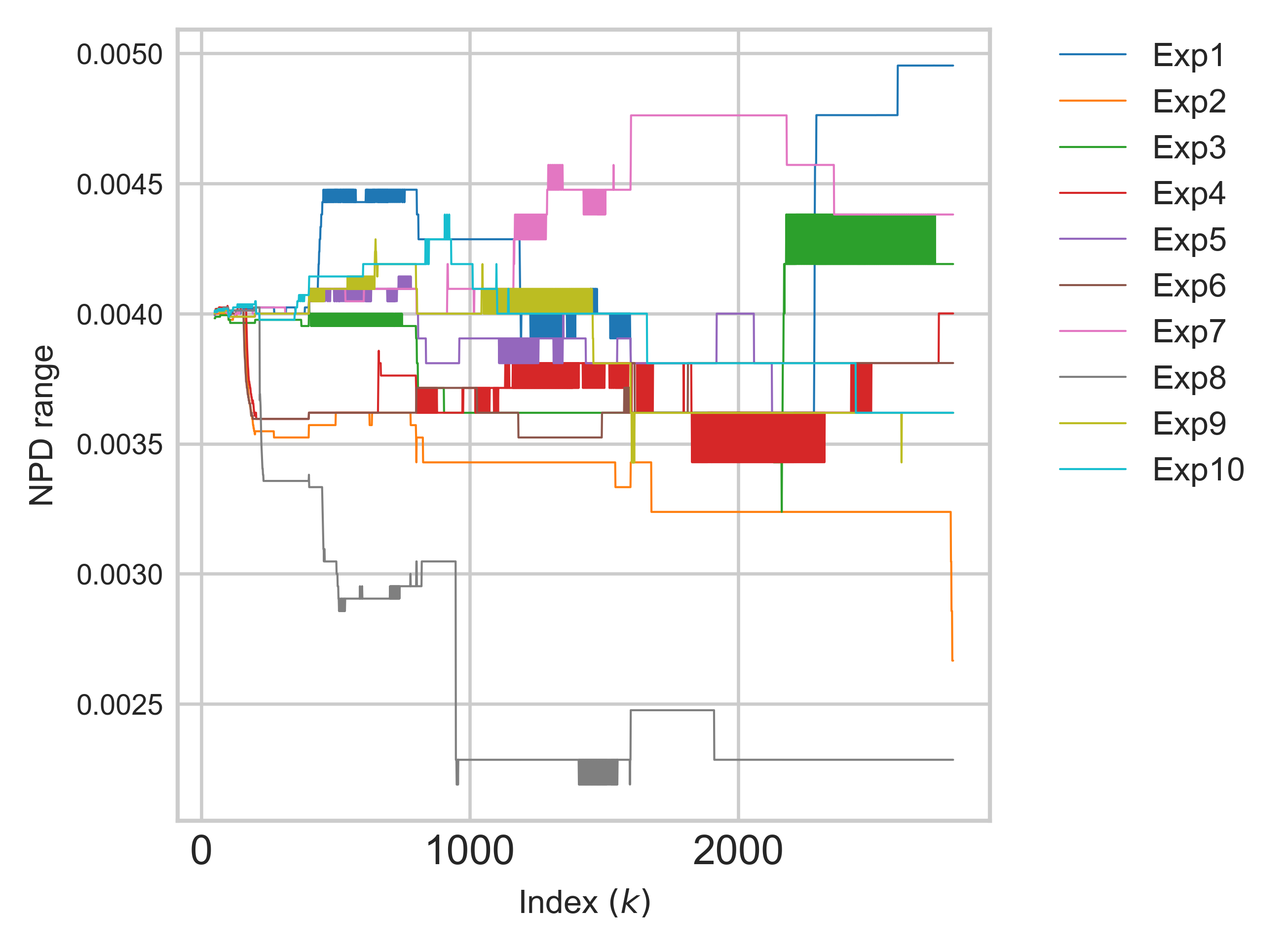}
    \vspace{-0.1cm}
		\caption{
		\ac{npd} range for $10$ experiments, with each mobile node moving at a different direction and traverses $1$ [Km] during the simulation.}
		\label{fig:NPD_walk_allexperimentsDiffDirectionsOFFLINE}
	\end{figure}

	Lastly, we examine  synchronization performance for the topology used in Fig. \ref{fig:FreqmoduloPhase_offlineTest}, with mobility, as was done in Fig. \ref{fig:NPD_walk_allexperimentsDiffDirections}, i.e., we randomly and uniformly select $30\%$ of the nodes, and each selected nodes a random angular direction is selected uniformly over $[0,2\pi)$. The moving nodes travel  at a fixed speed, such that at the end of the simulation, each moving node has travelled $1$ [Km]. 
	Fig. \ref{fig:NPD_walk_allexperimentsDiffDirectionsOFFLINE} depicts the evolution of \ac{npd} range with time, as was done in Fig. \ref{fig:NPD_walk_allexperimentsDiffDirections} for online training.
	Specifically, it was observed that  {\em at displacement of $1$~[Km], The \ac{npd} range increases by a factor smaller than $2$}, similar to online training. Thus, we conclude that {\em offline training also offers excellent robustness to node mobility.}
	
\color{black}
	\vspace{-0.4cm}
	\section{Conclusions}
	\label{sec:conclusion}
	\vspace{-0.3cm}
	This work considers network clock synchronization for wireless networks via pulse coupled \acp{pll} at the nodes. 
	The widely studied classic synchronization scheme based on the update rule \eqref{eqn:clock_analytical} is known 
	to fail in achieving full synchronization  for networks with non-negligible  propagation delays, and/or the clock frequency differences among the nodes,
	resulting in clusters of nodes synchronized among themselves, while the clocks of nodes belonging to different clusters are not phase-synchronized.
	In this work, we propose an algorithm, abbreviated as \ac{dasa}, which replaces the analytically computed $\aij$ coefficients   of the classic algorithm 	with weights learned  using \acp{dnn}, such that
    learning is done is an {\em unsupervised and distributed manner}, and requires {\em a very short training period}. These properties make the proposed algorithm very attractive for practical implementation.
	With the proposed \ac{dnn}-aided synchronization scheme, 
	each node determines its subsequent clock phase using its own clock and the timings of the pulses received from the other nodes in the network. 
	Numerical results show that when there are propagation delays and clock frequency differences between the nodes, both the proposed \ac{dasa} and the classic analytically-based scheme 
	achieve frequency synchronization, however {\em only the proposed \ac{dasa}   is able to attain full synchronization} of both the frequency and phase with a very high  accuracy. It was demonstrated that  \ac{dasa} maintains synchronization also in the presence of clock frequency and phase resets occurring at a subset of the nodes. Moreover, \ac{dasa} was also shown to maintain accurate synchronization  when only part of the nodes is mobile. Lastly we evaluated the relevance of offline training to the considered scenario: It was shown that offline training achieves full synchronization, with only a small degradation in the \ac{npd} and the \ac{npd} range, compared to online training.
	
	

\vspace{-0.1cm}

\bibliographystyle{IEEEtran}
\bibliography{IEEEabrv, myreference.bib}
	

\end{document}